\documentclass[journal,transmag]{IEEEtran}
\usepackage{cite}

\ifCLASSINFOpdf
   \usepackage[pdftex]{graphicx}
\else
   \usepackage[dvips]{graphicx}
\fi
\graphicspath{{./figures/}}
\usepackage{amsmath}
\ifCLASSOPTIONcompsoc
  \usepackage[font=normalsize,labelfont=sf,textfont=sf]{subfig}
\else
  \usepackage[font=footnotesize]{subfig}
\fi
\usepackage{url}

\usepackage{slashbox}

\begin{document}
%
\title{Analytical Derivation of Downlink Data Rate Distribution for 5G HetNets with Cell-Edge Located Small Cells}


\author{\IEEEauthorblockN{G\"uven Yenihayat,
Ezhan Kara\c{s}an}\\
\IEEEauthorblockA{Electrical and Electronics Engineering Department,
Bilkent University, Ankara, Turkey\\
Email:\{guven, ezhan\}@ee.bilkent.edu.tr}}

%




\maketitle

%
%
%
%
%
%
%

\begin{abstract}
In HetNets, time/frequency resources should be partitioned intelligently in order to minimize the interference among the users. In this paper, the probability distributions of per user downlink Data Rate, Spectral Efficiency (SE) and Energy Efficiency (EE) are analytically derived for a HetNet model with cell-edge located small cells. The high accuracy of analytically derived CDFs have been verified by the distributions obtained via simulations. CDF expressions have then been employed in order to optimize Key Parameter Indicators (KPI) which are selected here as $10^{th}$ percentile downlink user Data Rate ($R_{10}$), Spectral Efficiency ($SE_{10}$) and Energy Efficiency ($EE_{10}$). 

In addition to optimizing KPIs separately, employing the analytically derived distributions, we have also investigated the variation of the KPIs with respect to each other. The results have shown that the resource allocation parameter values maximizing $R_{10}$ is very close to the values that maximize $SE_{10}$. However, the values that are optimal for $SE_{10}$ and $R_{10}$, are not optimal for $EE_{10}$, which demonstrates the EE and SE trade-off in HetNets. We have also proposed a metric, $\theta$, aiming to jointly optimize SE and EE. The results have shown the value of resource sharing parameter optimizing $\theta$ is closer to the value that maximizes SE. This result shows that SE is more critical in SE-EE trade-off.
\end{abstract}


\section{Introduction}
\label{sec:Intro}
%
%
%
%


\IEEEPARstart{W}{ith} the ongoing evolution of mobile devices, the demand for higher data rates in mobile communication systems has been increasing exponentially. According to Wireless World Research Forum's (WWRF) vision for 2020, a mobile traffic growth of 1000 times is expected.  International Telecommunication Union (ITU) has opened a programme for this new cellular standard as International Mobile Telecommunication 2020 (IMT-2020) which will be the 5th generation of cellular standards (5G). According to 5G visions of ITU and several communication companies, the services in 5G will require larger data rates, lower latency and higher reliability. All the improvements should be done in a cost effective manner\cite{chin}. In order to satisfy the 1000x data challenge, as stated in \cite{WW5GBE} and \cite{BOCCARDI}, the key technologies are increased bandwidth, increased spectral efficiency and extreme cell densification.

Cell densification is a key enabler for 5G networks,  \cite{WW5GBE}, \cite{DAMNJANOVIC}. By shrinking the cell sizes, the spectrum can be reused across on the area and number of users using same resources will decrease which will increase the per user rate. In dense deployments, adding more BSs will also increase the interference levels. In order to overcome this problem, deployment of BSs with lower transmit power is proposed. Low power base stations are named as micro, pico, femto base stations depending on their transmit power. Networks consisting of a mix of these base stations are called Heterogeneous Networks (HetNets).

In HetNets, with the addition of small cells, area spectral efficiency is increased. For the ongoing 3GPP development, there are various scenarios and requirements for the enhancement of small cells \cite{GEN3PP}. Cell range expansion (CRE) is one of the methods in HetNets to offload more users to small cells. It is enabled by through cell biasing and adaptive resource partitioning and is seen as an effective method to balance the load among the nodes in the network and improve overall trunking efficiency. Depending on the bias value, the network can control the number of users (UE) associated with the low-power nodes and therefore control traffic demand at those nodes \cite{DAMNJANOVIC}. A positive bias means that the UE associate itself with small cell as long as the difference of the received power between macro cell and small cell is smaller than this bias value. In this study, we assume a two tier HetNet where there are Macro and Micro base stations and with cell biasing there are three types of UEs: the UEs that are associated with Macro cell are named as Macro UEs, the UEs associated with Micro cell with zero biasing are Direct Micro UEs and the UEs associated with Micro cell with positive biasing are named as Cell Range Expanded (CRE) UEs.

In order to minimize the interference among the users of the system, time/frequency resources should be partitioned carefully in HetNets. For instance, in case when CRE UEs and Macro UEs are being served in same time interval and at the same frequency band, the received signal power of CRE UEs will be lower than interference power coming from Macro cell. Therefore, the resources are needed to be orthogonally shared between CRE and Macro UEs. This can be done by using Almost Blank Subframe (ABS) technique which is a part of Enhanced Intercell Interference Coordination (eICIC) developed by 3GPP working group, \cite{Qiaoyang}. As stated in \cite{Qiaoyang} and \cite{SARABJOT_2}, in ABS Macro BS does not transmit in a portion of the time when CRE UEs are being served, so that CRE UEs do not suffer from the Macro BS interference.

The resource allocation between Direct Micro UEs and Macro UEs can be done by using orthogonal or non-orthogonal deployments. In \cite{ROSENBERG}, it is stated that the frequency band can be shared among Macro and Micro tier users by using one of three different schemes which are Co-channel deployment (CCD), Orthogonal deployment (OD) and Partially Shared deployment (PSD). In CCD all sub-channels can be used by Macro and Micro BSs whereas in OD the sub-channels are shared orthogonally among micro and macro tier users. PSD is a combination of CCD and OD. In PSD, a number of the sub-channels are used by both Macro and Micro BS whereas remaining sub-channels are used only by Macro BS.

In the literature, there are studies regarding how to partition the resources and adjust bias value for HetNets assuming different base station and user distributions, in order to maximize user rates. For instance, in \cite{SARABJOT}, rate distribution of users are found for a two tier HetNet in which, users and base stations are distributed on the area asuming a 2-D Poisson Point Distribution. Using the rate distribution, the authors analyzed the system for different bias and resource partitioning parameters setting $5^{th}$ percentile and median rate as performance indicators.

There are studies to obtain analytical downlink data rate distribution for HetNets and other types of networks. In \cite{GUVENC_1}, a semi-analytical approach depending on computer simulations is utilized to investigate the capacity of a HetNet scenario that is similar to our scenario. Using the semi-analytical approach for capacity distribution, the authors of \cite{GUVENC_1} optimize the HetNet system in terms of fairness and user data rates. Different than \cite{GUVENC_1}, in our study, we follow a fully analytical approach. We obtain an analytical expression for the user rate and do not need to do simulations for further optimization of system parameters. Additionally, in our system the bandwidth is shared among different types of users whereas in \cite{GUVENC_1}, co-channel deployment is assumed.

In many other studies in the literature, e.g., \cite{ANDREWS_1}, \cite{GANTI}, \cite{MUKHERJEE}, \cite{HEATH} the performance of HetNets are modeled using Poisson-Point-Process (PPP) based models since it provides computationally efficient methods for the analytical performance evaluations. However, as stated in \cite{Merwaday2014}, in PPP based models the macro base stations may be very close to each other which is not the case in real life scenarios. In a recent study given in \cite{Guo_Haenggi}, SINR distribution is investigated for a general class of point processes and it has been shown that SINR distribution obtained for a point process is a shifted version of the distribution obtained for the other point process. 

In this paper, different than the Hetnet model used in \cite{SARABJOT} and \cite{ANDREWS_1, GANTI, MUKHERJEE, HEATH, Guo_Haenggi}, we model a HetNet for a scenario in which small cells are located at cell edges which is known as Cell-On-Edge (COE) configuration. As stated in \cite{SHAKIR}, the COE configuration has been shown to produce significant gains to operators and mobile customers, including improved cell-edge coverage, increased network capacity to match cell-edge mobile user demands, enhanced end-user experience, and reduced cost of delivering mobile broadband services to such cell-edge mobile users.

According to Next Generation Mobile Networks (NGMN) alliance, in next generation communication systems energy efficiency of the networks is a key factor and it is a central design principle of 5G. Energy efficiency is defined as the number of bits that can be transmitted per Joule of energy, where the energy is computed over the whole network. As stated above 5G should support a 1,000 times traffic increase, but this increase should be done with an energy consumption by the whole network of only half that typically consumed by today’s networks. This leads to the requirement of an energy efficiency increase of x2000 \cite{NGMN}. One way to reduce power consumption is bandwidth expansion. In \cite{SE_EE_TOFF_WU}, it is stated that bandwidth expansion enables savings of power consumption of up to 45\% if all cells in the network apply the same bandwidth expansion strategy. In this study, in addition to user rate analysis we have also investigated the variation of Spectral and Energy Efficiency with resource allocation parameters in HetNet model under consideration. 

In this study, for the HetNet model with Cell-Edge Located Small Cells, we use a resource allocation scheme which partitions the resources in time and also in frequency domain. We analytically derive the probability distribution of downlink data rates achieved by users in the cell and then verify the proposed analytical model by simulations. We show that the distributions given by the analytical model are highly accurate under different network parameters such as user distribution and bias. By using the analytical rate distribution expression, we optimize the system in terms of $10^{th}$ percentile rate ($R_{10}$). Additionally, by using a similar approach that we have followed to obtain rate distribution, we have also derived the CDF expressions for Spectral Efficiency (SE) and Energy Efficiency (EE). We have also optimized the resource allocation parameters in order to maximize tenth percentile SE ($SE_{10}$) and EE ($EE_{10}$). These results demonstrate the SE and EE trade-off in the studied HetNet model.

The most important contributions of this paper are:

\begin{itemize}
	\item Analytical CDF of downlink data rate per user, SE and EE are derived for a HetNet model with COE configuration, for different user distributions and bias values.
	\item The analytically derived CDFs have been shown to be very close to CDFs obtained by simulations.
	\item Using analytically obtained distributions, optimal resource allocation parameters have been obtained that maximize $R_{10}$, $SE_{10}$ and $EE_{10}$.
	\item The results have shown that the optimal values of resource allocation parameters maximizing $R_{10}$ and $EE_{10}$ are close to each other, however these values are not optimal for $SE_{10}$.
	\item A single performance metric for jointly maximizing SE and EE has been proposed and the variation of this parameter with HetNet system parameters has also been analyzed. The results have shown the value of resource allocation parameter optimizing joint metric is closer to the value that maximizes SE. 
\end{itemize}

The rest of this paper is organized as follows: Section \ref{sec:System_Model} describes our system model for the HetNet in consideration. Section \ref{sec:Analitic} presents the derivation of cumulative distribution function of the rate/user, spectral efficiency and energy efficiency for the network model and resource allocation scheme employed. Section \ref{sec:Results} presents the simulation and analytical results and the paper is concluded in Section \ref{sec:Conclusion}.

\section{System Model }
\label{sec:System_Model}

We consider a heterogeneous network model that consists of Macro and Micro Base Stations (BS) and User Equipments (UE). We assumed the Cell-On-Edge Model in which Micro BSs are located on the edge of a cell that is covered by a Macro BS.  The model follows the assumptions, which are in accordance with the 3GPP simulation parameters given in \cite{GEN3PP_2}, are enlisted below;
\begin{itemize}
	\item There is 1 Macro BS located in the center of a circular area with a radius of 1km.
	\item There are 10 Micro BSs which are located on the ring that is located 0.8 km away from the center. The distance between adjacent Micro BSs are equal.
	\item A part of the user equipments (UEs) are located uniformly on all the area.
	\item The other part  of the UEs are located uniformly within circles that are in the coverage of Micro BSs. The ratio of this part of UEs to all UEs located on all area is equal to $W_{Micro}$.
\end{itemize}

\begin{figure}
	\includegraphics[scale=0.6]{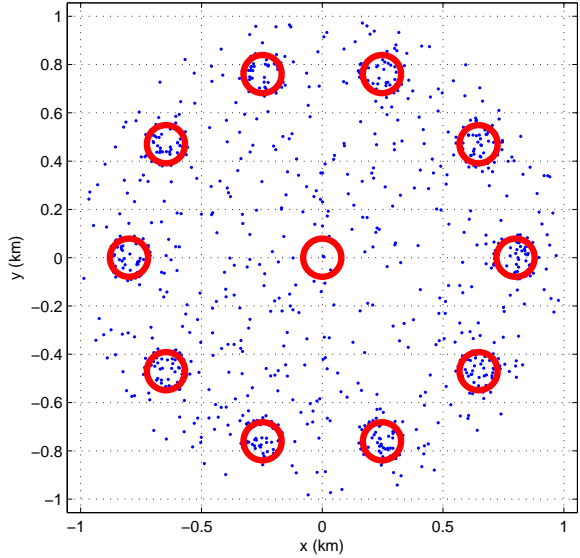}
		\caption{An example of topology, red circles are Base Station Locations} \label{fig:exTop}
\end{figure}

One example topology is shown in Figure \ref{fig:exTop}.  By examining Figure \ref{fig:exTop}, the denser distribution of the users around micro base stations can be observed. As $W_{Micro}$ increases, the density around Micro BS will also increase.

In the communication system model, we have only considered the down-link communication from the BSs to UEs and assumed that UEs have always something to receive from BSs. The wireless channel between BSs and UEs is modeled by a path loss model for which the received power ($P_{r,i}$ in watts) is related to transmit power ($P_{t,i}$ in watts) of BS with index $i$ as in

\begin{equation}
P_{r,i}= \frac {P_{t,i}} {d_i ^ {\gamma_i} } \label{eqn:P_rec_i}.
\end{equation}

Throughout the paper we use the following convention for BS type to index mapping: the BS with index $i=0$ is the Macro BS and the BSs with index $i>0$ are Micro BSs. In (\ref{eqn:P_rec_i}), $d_i$ is the distance between UE and BS $i$ and $\gamma_i$ is the attenuation parameter for BS $i$ and it is defined by

\begin{equation}
	\gamma_i=\begin{cases} \alpha_1, & \mbox{if } i \mbox{=0} \\ \alpha_2, & \mbox{if } i \mbox{$>$0} \end{cases}  \label{eqn:alpha_vals}.
\end{equation}

In (\ref{eqn:P_rec_i}), $P_{t,i}$ differs depending on the BS type and it is given as

\begin{equation}
	P_{t,i}=\begin{cases} P_1, & \mbox{if } i \mbox{=0} \\ P_2, & \mbox{if } i \mbox{$>$0} \end{cases}.  \label{eqn:Pt_vals}
\end{equation}

In this system, each UE calculates its signal power parameter $P_{s,i}$, which is a scaled version of received power $P_{r,i}$ with the bias value of BS $i$ ($B_i$). Depending on the $P_{s,i}$ value, UE associates itself with a BS. The relation between $P_{s,i}$ and $P_{r,i}$ is given by  

\begin{equation}
P_{s,i}=P_{r,i} 10^{ \frac  {B_i} {10} } \label{eqn:P_s_i}
\end{equation}

where

\begin{equation}
B_i=\begin{cases} 0, & \mbox{if } i \mbox{=0} \\ B, & \mbox{if } i \mbox{$>$0} \end{cases}  \label{eqn:B_vals}.
\end{equation}

\begin{figure}
	\includegraphics[scale=0.6]{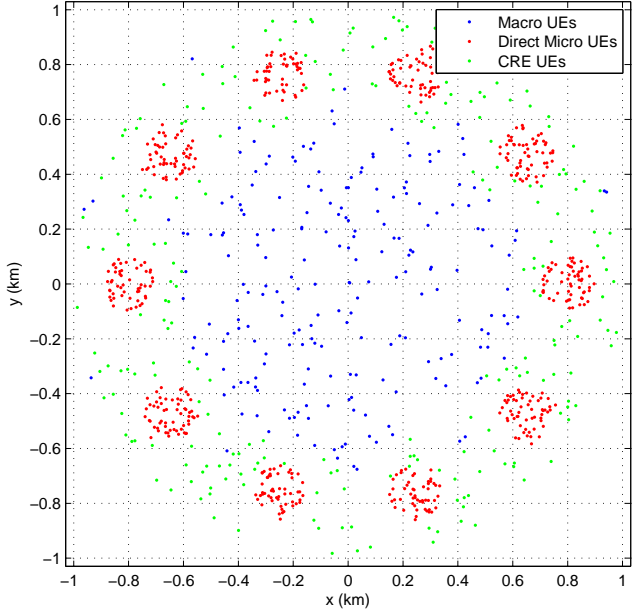}
	\caption{UE Connections for B=15dB} \label{fig:Top_B_15}
\end{figure}

Each UE is associated with the BS $i$ for which signal power parameter, $P_{s,i}$ is maximum. In the system each UE can be a Macro, Cell Range Extended (CRE) or a Direct Micro UE. The UEs for which $P_{s,i}$ is maximum for $i=0$ are Macro UEs, the UEs for which $P_{s,i}$ is maximum for $i>0$ and $B=0$ are Direct Micro UEs, and the UEs for which $P_{s,i}$ is maximum for $i>0$ and $B>0$ are CRE UEs. Figure \ref{fig:Top_B_15} shows how UEs are associated with BSs for $B=15$ dB.

\begin{figure}
	\includegraphics[scale=0.5]{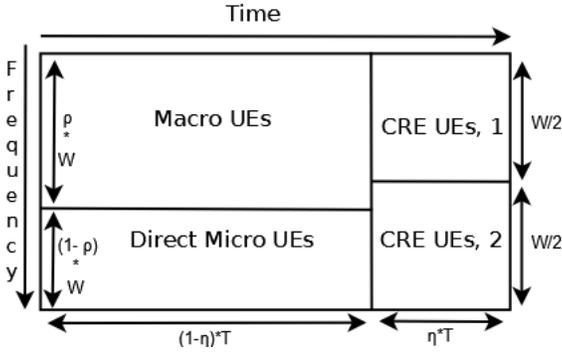}
	\caption{Allocation of time and frequency to UEs} \label{fig:case_4}
\end{figure}

In our system we use a resource allocation scheme which is shown by Figure \ref{fig:case_4}. According to this resource allocation scheme, CRE UEs are served in $\eta$ of time for $0 \leq \eta \leq 1$ whereas Direct Micro UEs and Macro UEs are served in $1-\eta$ amount of time. In addition to partitioning in time domain, we have also employed a partitioning in frequency domain. The whole band is shared orthogonally among adjacent Micro BSs in order to reduce interference coming to CRE UEs from neighboring Micro BSs. Frequency band is orthogonally allocated among Macro and Direct Micro UEs, so that the interference power at Direct Micro UEs and Macro UEs is minimized. In our scheme, $\rho W$ of the total band $W$ will be used by Macro UEs.


The time/frequency resources that are given to UEs are shared equally among UEs that are connected to the same BS. For example, if a Micro BS has $N_m$ Direct Micro UEs, each user has a total access to a channel having a bandwidth of $(1-\rho)W$ for $\frac{1-\eta}{N_m}$ in one unit of time.

We have also assumed that each UE uses the maximum capacity of the channel assuming Gaussian alphabet is transmitted.

Using the communication model described above, we have investigated parameters such as the cumulative distribution (CDF) of bit rate per UE, spectral and energy efficiency (SE and EE). We have selected $10^{th}$ percentile rate ($R_{10}$), median and tenth percentile Spectral ($SE_{50}, SE_{10}$) and Energy Efficiency ($EE_{50}, EE_{10}$) as Key Parameter Indicators (KPIs). We analytically derived CDF of bit rate per UE using a geometrical approach and verified our analytical results using extensive simulations. A similar approach has also been followed to derive the analytical distributions of SE and EE. Moreover, we have employed the analytical CDF expression of rate, SE and EE in order to select optimal values of resource sharing parameters ($\eta$, $\rho$) for two specific bias values $B$ in order to maximize the KPIs.

\section{Analytical Derivation of Cumulative Distribution of User Rate, Spectral and Energy Efficiency}
\label{sec:Analitic}

In this section, first the analytical formulas for cumulative distribution of capacity/user for the UEs in the heterogeneous network model will be derived by using a geometrical approach. Then, spectral and energy efficiency will be introduced and following a similar procedure the cumulative distributions of these two parameters will also be derived.

As described in Section \ref{sec:System_Model}, there are three types of UEs in the system, which are Macro, Direct Micro and CRE UEs. Each type of these users have different capacity distributions and for given $B$, $\eta$, $\rho$ values a general equation for capacity per user is given as,

\begin{equation}
C_{\zeta}(B, \eta, \rho)=\frac{\eta_\zeta}{N_{BS, \zeta}(B)} W \rho_\zeta log_2(1+ \frac{P_r}{\sigma^2_{N+I, \zeta}}). \label{eqn:Cap_UE}
\end{equation}

In (\ref{eqn:Cap_UE}), $\zeta$ is the index showing type of user. $\eta_\zeta$ is the time sharing parameter for user type $\zeta$. $N_{BS, \zeta}(B)$ is the average number of type $\zeta$ users per base station for bias value $B$, $W$ is the total bandwidth used, $\rho_\zeta$ is the band sharing parameter for user type $\zeta$. $P_r$ is the power of signal received from the associated BS. $\sigma^2_{N+I, \zeta}$ is the variance of $Interference + Noise$ term of user type $\zeta$, where this term is modeled as a Gaussian random variable. $\zeta$, $\eta_\zeta$ and $\rho_\zeta$ are given as,

\begin{equation}
\zeta=\begin{cases} 0, & \mbox{For a macro user} 
\\1, & \mbox{For a direct user}
\\  2, & \mbox{For a CRE user}, \end{cases}
\label{eqn:zeta_types}
\end{equation}

\begin{equation}
\eta_\zeta=\begin{cases} 1-\eta, & \mbox{$\zeta=0, 1$} 
\\\eta, & \mbox{$\zeta=2$}, \end{cases}
\label{eqn:eta_types}
\end{equation}

\begin{equation}
\rho_\zeta=\begin{cases} \rho, & \mbox{$\zeta=0$} 
\\1-\rho, & \mbox{$\zeta=1$}
\\0.5, & \mbox{$\zeta=2$}. \end{cases}
\label{eqn:rho_types}
\end{equation}

The parameters $\eta$, $\rho$ above, are time and band sharing parameters which are defined as shown in resource allocation scheme illustrated in Fig. \ref{fig:case_4}. In this paper, after obtaining the analytical expression for capacity distribution, we aim to find optimum values of $\eta$ and $\rho$ in order to maximize tenth percentile rate, $R_{10}$.

\subsection{Modeling $Interference+Noise $ Term}

$Interference+Noise $ term is modeled as a zero-mean Gaussian random variable with variance $\sigma^2_{N+I, \zeta}$. Due to the symmetry of the base station locations in heterogeneous network model, $\sigma^2_{N+I, \zeta}$ is assumed to be same for all users of same type. In this section, the variance value for different type of UEs will be presented.

\textbf{Direct Micro UEs:} Sources of interference for Direct Micro UEs are all micro base stations other than the one that is associated with the UE. We have assumed that the total $Interference+Noise$ for these users can be modeled as a Gaussian random variable with variance given by

\begin{equation}
\sigma^2_{N+I, 1}= P_{noise} + \displaystyle \sum _{i=2}^{N_{MICRO}} \frac {P_2} {l_i ^ { \alpha_2 } } \label{eqn:sigma_dir},
\end{equation}
where $l_i$ is the distance between $i^{th}$ Micro BS and $1^{st}$ Micro Base Station for $i=2,3,...N_{MICRO}$.

\textbf{CRE UEs:} Sources of interference for CRE UEs are the micro base stations that use the same portion of the band. Because the band is orthogonally and equally shared by Micro BSs. We have assumed that the total $Interference+Noise$ can be modeled as a Gaussian random variable with variance

\begin{equation}
\sigma^2_{N+I, 2}= P_{noise} + \displaystyle \sum _{i=2n+1}^{N_{MICRO}} \frac {P_2} {l_i ^ { \alpha_2 } } \label{eqn:sigma_CRE},
\end{equation} where $l_i$ is as defined above.

\textbf{Macro UEs:} By inspecting the resource allocation scheme shown in Fig \ref{fig:case_4}, it can be observed that there is no source of interference for Macro UEs, therefore $Interference + Noise$ term depends only on noise, therefore $\sigma^2_{N+I, 0}$ is given by 

\begin{equation}
\sigma^2_{N+I, 0}= P_{noise} \label{eqn:sigma_macro}.
\end{equation}

\subsection{Distribution of Received Power}
\label{sec:Dist_general}
\begin{figure*}
	\centering
	\includegraphics[width=17cm,height=9cm]{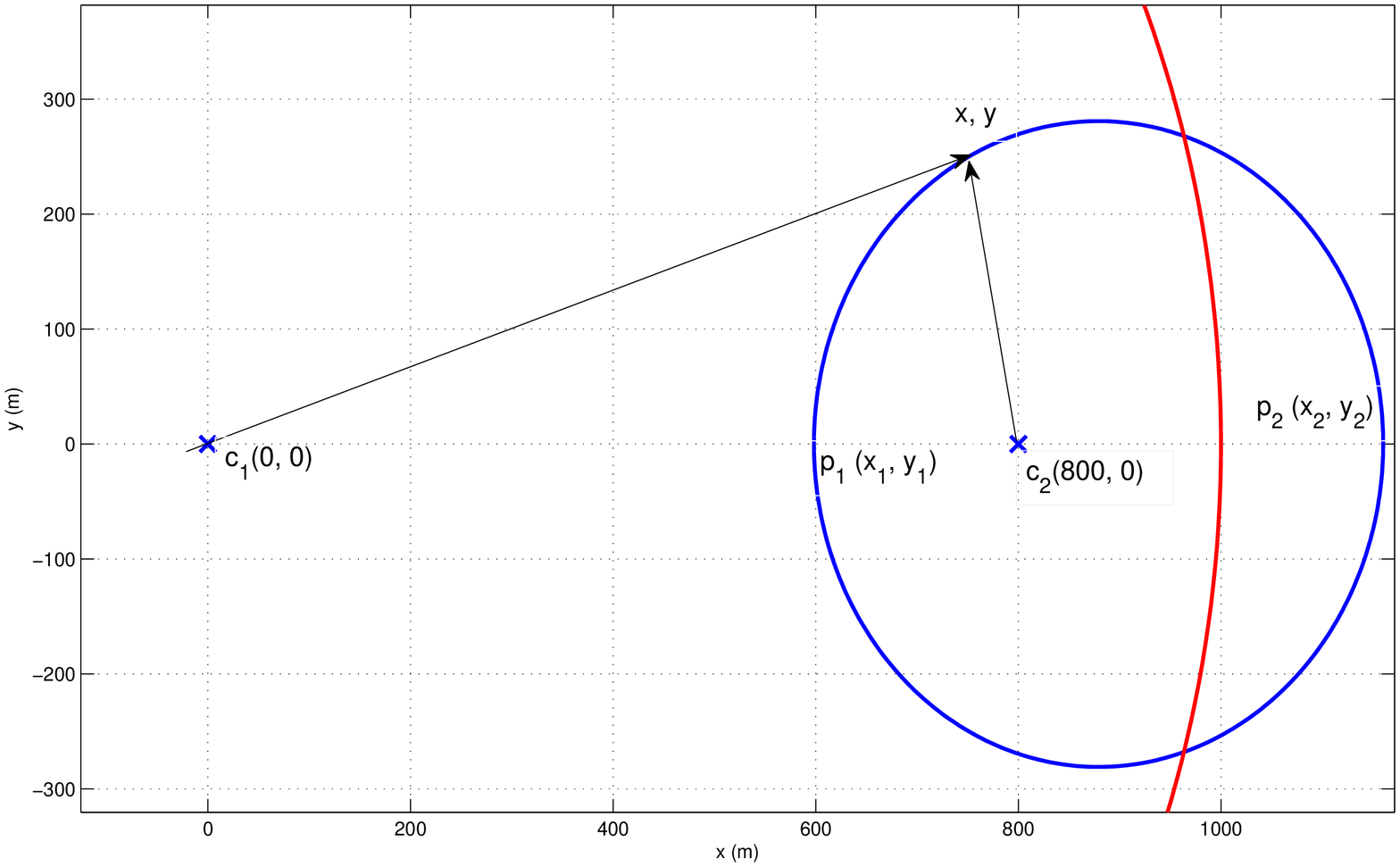}
	\caption{CRE region contour for $B=15dB$}
	\label{fig:CRE_Contour_15_dB}
\end{figure*}

With the Gaussian assumption for $Interference + Noise$ term, the distribution of received power, $P_r$, should be obtained first in order to find the distribution of capacity per user which was given by (\ref{eqn:Cap_UE}). Figure \ref{fig:CRE_Contour_15_dB} shows the range extended coverage of Micro BS for $B=15dB$. This is the region for which $P_{r, Micro} 10^{\frac{B}{10}} > P_{r, Macro}$, where $P_{r, Micro}$ is the power received from the closest Micro BS, $P_{r, Macro}$ is the received power from Macro BS and $B$ is the BIAS parameter.

Any point $x, y$ on the contour of this area should satisfy

\begin{equation}
\frac {P_1} {(x^2+y^2) ^ {\frac {\gamma_1} {2} }} = 10^{\frac {B} {10}} \frac {P_2} {((x-800)^2+y^2) ^ {\frac {\gamma_2} {2} } }.  \label{eqn:Contour_eqn}
\end{equation}

\begin{figure}
	\includegraphics[scale=0.5]{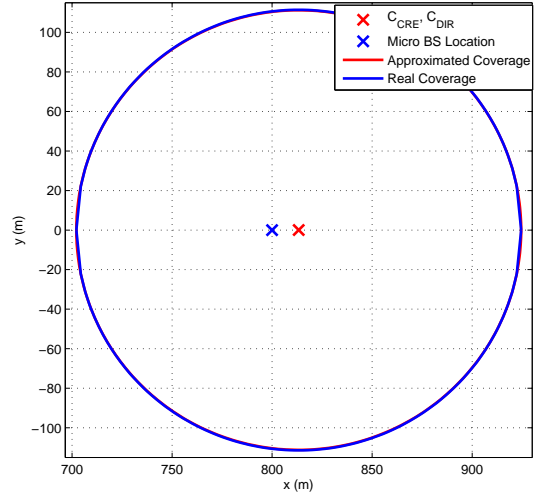}
	\caption{Approximated coverage for $B=0dB$}
	\label{fig:Coverage_0_dB}
\end{figure}

\begin{figure}
	\includegraphics[scale=0.5]{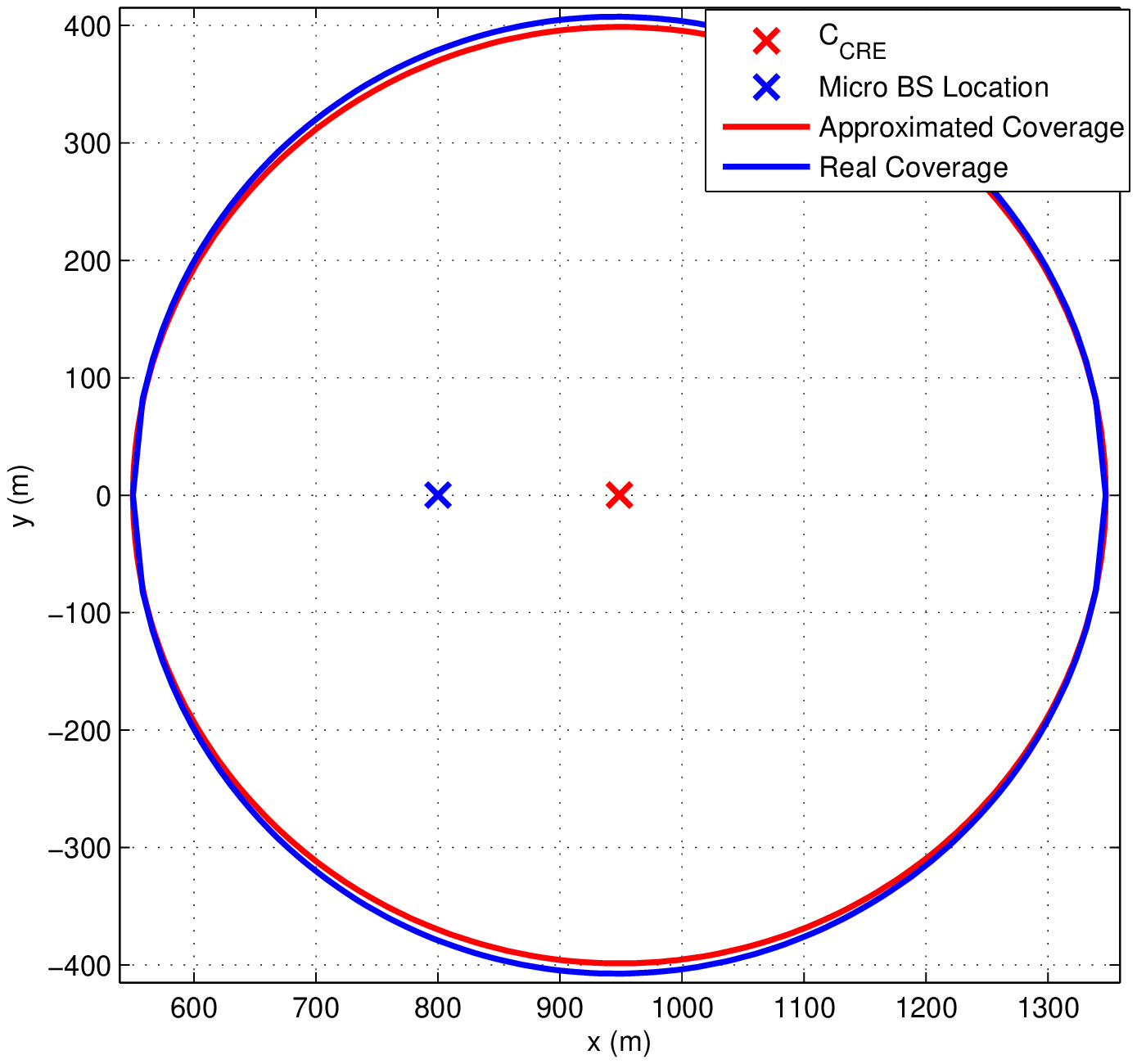}
	\caption{Approximated coverage for $B=20dB$}
	\label{fig:Coverage_20_dB}
\end{figure}

\begin{figure}
	\includegraphics[width=9.0cm,height=8.5cm]{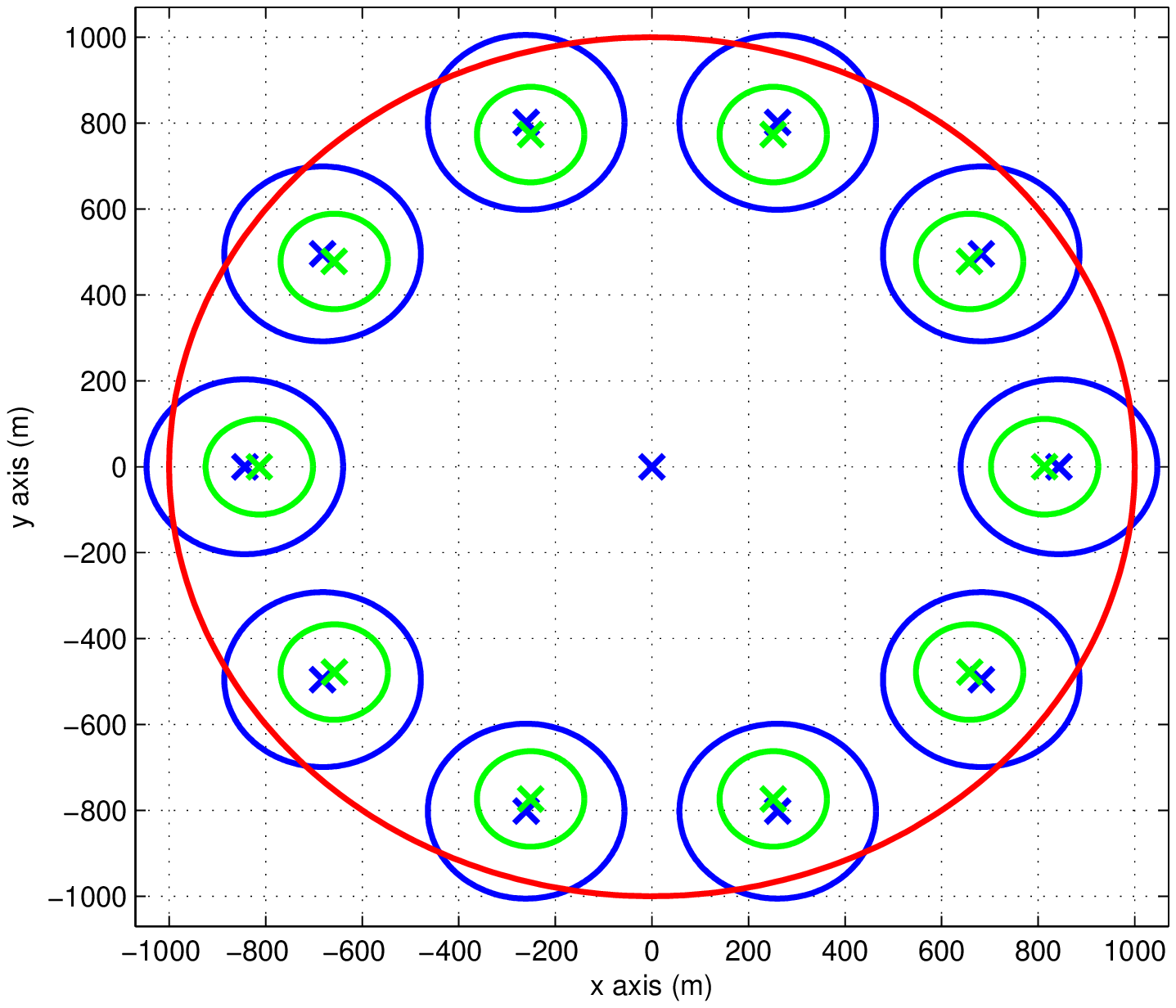}
	\caption{Coverage of different type of BSs for $B=10dB$}
	\label{fig:AllArea_10_dB}
\end{figure}

\begin{figure}
	\includegraphics[width=9.0cm,height=8.5cm]{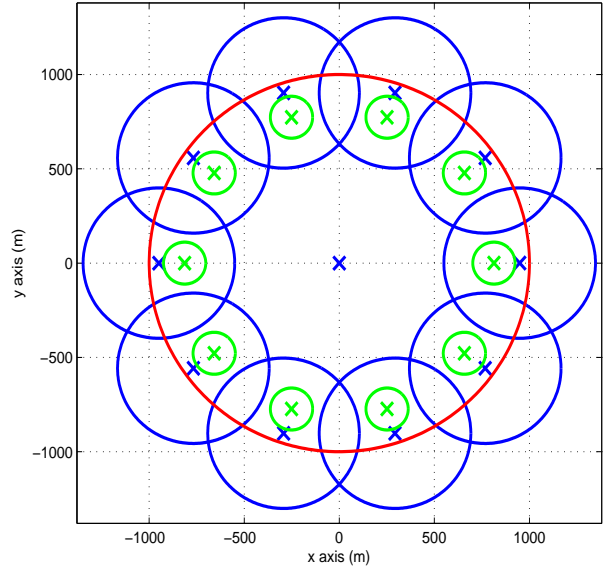}
	\caption{Coverage of different type of BSs for $B=20dB$}
	\label{fig:AllArea_20_dB}
\end{figure}

(\ref{eqn:Contour_eqn}) is numerically solved for various $B$ values and coverage of Micro BS is obtained. In order to simplify the analytical calculations, these coverage regions are approximated by circles centered at  points $c=\frac {p_1+p_2} {2} e ^ {j \frac{\pi i}{N_{MICRO}}}, \: i=0,1,2 ...(N_{MICRO}-1) $ and have the radius of $r=\frac {|p_1-p_2|} {2}$. $p_1, p_2$, which are shown in figure \ref{fig:CRE_Contour_15_dB}, are points where the contour intersects the $y=0$ line, and $|\cdot|$ is the magnitude operator. Figures \ref{fig:Coverage_0_dB} and \ref{fig:Coverage_20_dB} show how this approximation is close to the actual coverage region for two different bias values, $B=10dB$ and $B=20dB$. The approximated coverage model of the BSs for same bias values are shown in Figures \ref{fig:AllArea_10_dB} and \ref{fig:AllArea_20_dB}, respectively.

Using this approximated model for the system, the cumulative distribution of received power ($P_r$) for different type of users will be derived in sections \ref{sec:Pr_Dir_Micro}, \ref{sec:Pr_CRE} and \ref{sec:Pr_Macro}. Using CDFs obtained for $P_r$ for different type of UEs, the capacity per any UE will be derived in section \ref{sec:Cap_User}.

\subsubsection{Distribution of Received Power for Direct Micro Users}
	\label{sec:Pr_Dir_Micro}
	
	Received power for a UE is a function of the distance ($D$) between the UE and the BS that is associated with it. Therefore, in order to obtain received power distribution, first we have to obtain CDF of random variable $D$, i.e. $F_D(d)=P(D \le d)$. As shown in Figure \ref{fig:Coverage_0_dB}, direct micro coverage is approximated by a perfect circle with parameters defined in Section \ref{sec:Dist_general}. Using this approximation and keeping uniform user distribution in mind, $F_D(d)$ for a direct micro user can be calculated as 
	
	\begin{equation}
	P(D \le d) = \frac {S(d)} {S_{DIR}},  \label{eqn:CDF_Micro}.
	\end{equation}
	
\begin{figure}
		\includegraphics[scale=0.25]{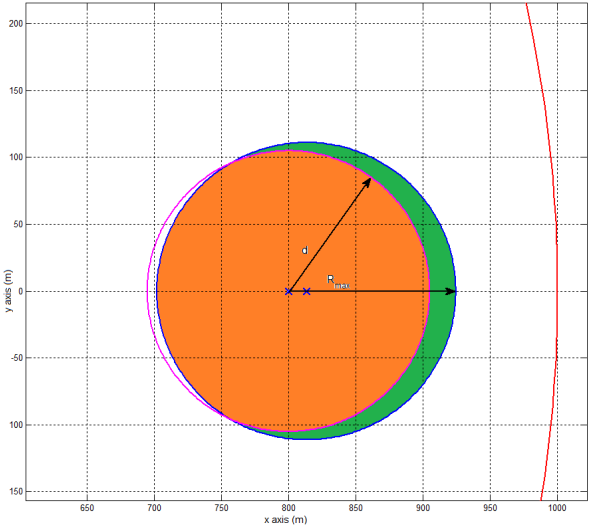}
		\caption{Calculation of Direct Micro Users' distance distribution to serving Micro BS}
		\label{fig:DIR_Region_plot}
	\end{figure}

	In (\ref{eqn:CDF_Micro}), $S(d)$ is the intersection area of the circle centered at Micro BS location with a radius of $d$ ($d \leq R_{max}$) and approximated coverage region for Micro BS (orange colored region in Figure \ref{fig:DIR_Region_plot}). $R_{max}$ is the maximum distance between micro BS location and a Direct Micro UE. $S_{DIR}$ is the area of the approximated direct Micro coverage region (union of orange and green colored regions in Fig. \ref{fig:DIR_Region_plot}). Calculation of the intersection area of two circles is given in Appendix \ref{sec:TwoCircIntersection}.
	
	Using (\ref{eqn:CDF_Micro}) and the relation between distance and received power that is given by (\ref{eqn:P_rec_i}), the CDF of received power $P_R$ can be expressed as
	
	\begin{equation}
	F_{p_r}(P_R) = 1-F_D(\sqrt[-\alpha_2] {\frac{P_R \sigma^2_{N+I, 1}}{P_2}}). \label{eqn:Pr_DirMicro_CDF}
	\end{equation}

\subsubsection{Distribution of Received Power for CRE UEs}
	\label{sec:Pr_CRE}
	
\begin{figure}
		\includegraphics[scale=0.3]{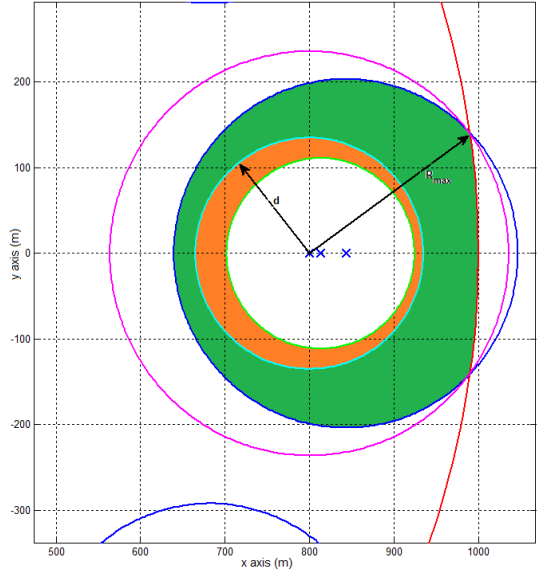}
		\caption{CRE Region in detail for $B=10dB$}
		\label{fig:CRE_Detail_10_dB}
	\end{figure}

\begin{figure}
		\centering
		\includegraphics[scale=0.4]{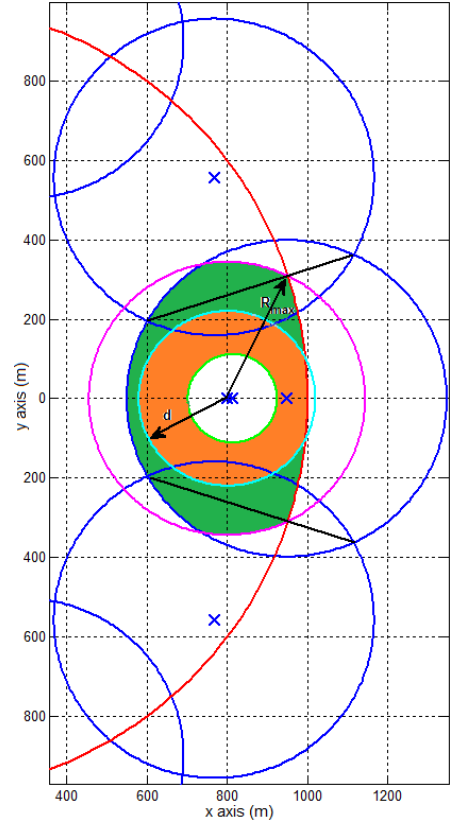}
		\caption{CRE Region in detail for $B=20dB$}
		\label{fig:CRE_Detail_20_dB}
	\end{figure}

	The distribution of received power for CRE UEs can be calculated similarly to that of Direct Micro UEs. The approximated coverage model for $B=10dB$ and $B=20dB$ are shown in Figures \ref{fig:CRE_Detail_10_dB}-\ref{fig:CRE_Detail_20_dB}. The range extended coverage of Micro BS is approximated by a perfect circle as defined in Chapter \ref{sec:Dist_general}. Using uniform distribution of UEs, the distribution of the distance between CRE UEs and Micro BS is given by
	
	\begin{equation}
	P(D \le d) = \frac {S(d)} {S_{CRE}(B)}  \label{eqn:CDF_CRE}.
	\end{equation}
	
	In (\ref{eqn:CDF_CRE}), $S(d)$ is the area of the intersection of the circle centered at Micro BS location with a radius $d$ ($d \leq R_{max}$), with approximated CRE region and all region where UEs are uniformly located. This region is colored to orange in Figures \ref{fig:CRE_Detail_10_dB} and \ref{fig:CRE_Detail_20_dB} for different $B$ values. $S_{CRE}(B)$ is the total area of region where CRE users are located for a fixed bias value $B$, geometrically this area can be expressed by the union of orange and green colored regions that are shown in Figures \ref{fig:CRE_Detail_10_dB} and \ref{fig:CRE_Detail_20_dB}. 
	
	Using (\ref{eqn:CDF_CRE}) and (\ref{eqn:P_rec_i}), the CDF of received power, $P_R$ can be written as

	\begin{equation}
	F_{p_r}(P_R) = 1-F_D(\sqrt[-\alpha_2] {\frac{P_R \sigma^2_{N+I, 2}}{P_2}}) \label{eqn:Pr_CRE_CDF}.
	\end{equation}
	

\subsubsection{Distribution of Received Power for Macro UEs}
	\label{sec:Pr_Macro}

\begin{figure}
		\includegraphics[scale=0.35]{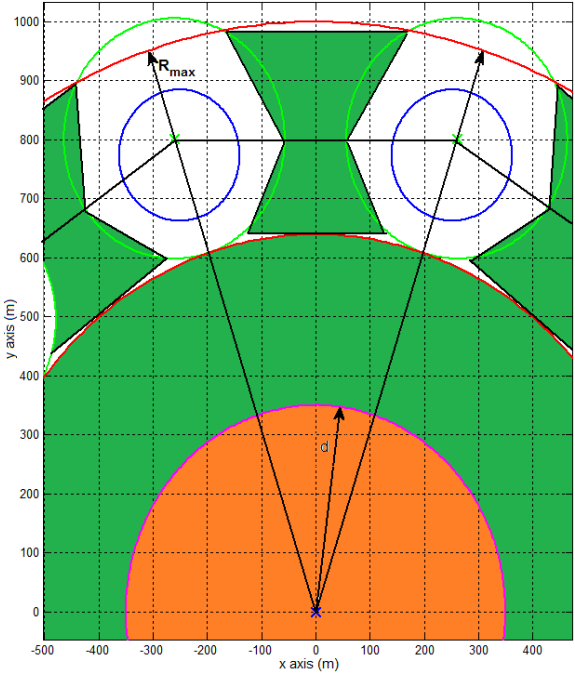}
		\caption{Macro Region in detail for $B=10dB$}
		\label{fig:MACRO_Detail_10_dB}
	\end{figure}

\begin{figure}
		\includegraphics[scale=0.4]{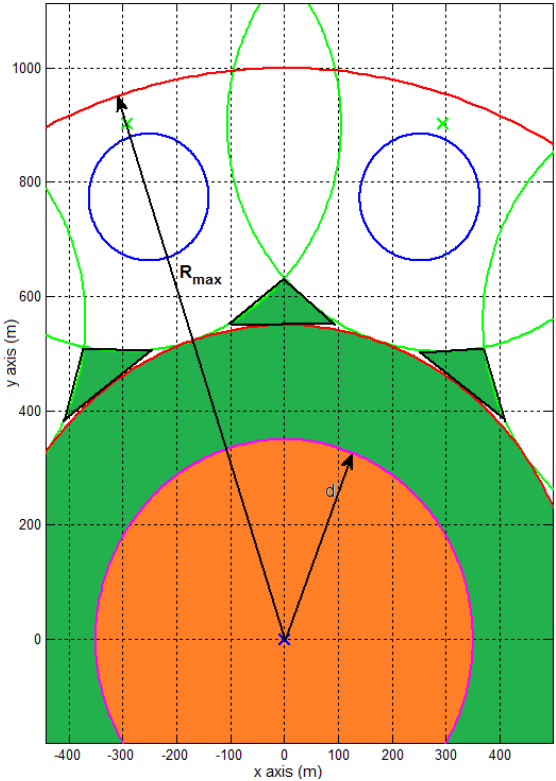}
		\caption{Macro Region in detail for $B=20dB$}
		\label{fig:MACRO_Detail_20_dB}
	\end{figure}

	Distribution of the received power for Macro UEs can be found by calculating the distribution of the distance between Macro UEs and Macro BS. The Macro BS coverage region is modeled as a combination of differently shaped regions as illustrated in Figures \ref{fig:MACRO_Detail_10_dB} and \ref{fig:MACRO_Detail_20_dB} for different $B$ values. The CDF of the distance between Macro UEs and Macro BS is given by

%
%
	
	\begin{equation}
		P(D \le d) = \frac {S(d)} {S_{MACRO}(B)}  \label{eqn:CDF_MACRO}.
	\end{equation}
	
	In (\ref{eqn:CDF_MACRO}), $S(d)$ is the area of the intersection of the circle centered at Micro BS location with a radius $d$  ($d \leq R_{max}$), with approximated Macro region. This region is colored to orange in Figures \ref{fig:MACRO_Detail_10_dB} and \ref{fig:MACRO_Detail_20_dB}. $S_{MACRO}(B)$ is the total area of the region where Macro users are uniformly located for a given bias value $B$, geometrically it is the union of orange and green colored regions in Figures \ref{fig:MACRO_Detail_10_dB} and \ref{fig:MACRO_Detail_20_dB}. 
	
	Using (\ref{eqn:CDF_MACRO}) and (\ref{eqn:P_rec_i}), the CDF of received power $P_r$ is given by 
	
	\begin{equation}
		F_{p_r}(P_R) = 1-F_D(\sqrt[-\alpha_1] {\frac{P_R \sigma^2_{N+I, 0}}{P_1}}) \label{eqn:Pr_Macro_CDF}.
	\end{equation}

\subsection{Distribution of Capacity Per User}
	\label{sec:Cap_User}

	The relation between the capacity per user $C$ and $P_r$ was given in (\ref{eqn:Cap_UE}). Equations (\ref{eqn:Pr_DirMicro_CDF}), (\ref{eqn:Pr_CRE_CDF}) and (\ref{eqn:Pr_Macro_CDF}) give the CDF's of $P_r$ for each type of UEs. Using these equations and (\ref{eqn:Cap_UE}) the capacity distribution for UE type $\zeta$can be obtained as in (\ref{eqn:C_Zeta}) in terms of the CDF of $P_r$.

	\begin{equation}	
		F_{C}(c|\zeta, B, \eta, \rho) = F_{Pr}(\sigma^2_{N+I, \zeta} (2^{\frac {c N_{BS, \zeta}(B)} {\eta_{\zeta}W\rho_{\zeta}} }-1)), \label{eqn:C_Zeta}
	\end{equation}
	
	Using (\ref{eqn:C_Zeta}), the distribution of capacity per any UEin the system can be obtained as given by
	
	\begin{equation}
		F_{C}(c|B, \eta, \rho) = \displaystyle \sum _{\zeta=1}^{3} P(\zeta|B) F_{C}(c|\zeta, B, \eta, \rho) \label{eqn:Cap_Any}
	\end{equation}
	where $P(\zeta|B)$ for a given $B$ value is given by 
	
	\begin{equation}
		P(\zeta=i|B) = \frac {  N_{\zeta}(B) } {N}, \zeta=0,1,2. \label{eqn:P_UE_Type}
	\end{equation}

	In (\ref{eqn:P_UE_Type}), $N_{\zeta}(B)$ is the average number of UEs of type $\zeta$ for bias value $B$ and $N$ is the total number of UEs in the system. According to system model, a ratio of $1-W_{Micro}$ of all UEs are distributed uniformly to all area, and a ratio of $W_{Micro}$ of UEs are distributed uniformly in Direct Micro coverage area. Using this model $N_{\zeta}(B)$ is given by 
	
	\begin{equation}
		E  N_{ \zeta }(B) =\begin{cases}
		N  (1-W_{micro}) \frac {S_{MACRO}(B)} {S_{TOT}} , & \mbox{$\zeta=0$}  \\
		N  W_{micro} + N  (1-W_{micro}) \frac {S_{DIR}} {S_{TOT}}, & \mbox{$\zeta=1$} \\
		N  (1-W_{micro}) \frac {S_{CRE}(B)} {S_{TOT}} & \mbox{$\zeta=2$}
		\end{cases}  \label{eqn:N_zeta}.
	\end{equation}	
	
	In order to calculate the areas of different type of UEs, calculation of intersection area of three circles should be calculated for large $B$ values. The method used for this calculation is shown in Appendix \ref{sec:ThreeCircIntersection}.
	
%
%
\subsection{Distribution of Spectral and Energy Efficiency}
\label{sec:EE_SE_Section}

Spectral Efficiency (SE) is defined as the experienced data rate of a UE per bandwidth occupied by the UE. SE is expressed by 

\begin{equation}
SE_{\zeta}(\eta)=\eta_\zeta log_2(1+ \frac{P_r}{\sigma^2_{N+I, \zeta}}). \label{eqn:SE_UE}
\end{equation}

Energy Efficiency is the rate of UE divided by the total power consumed by the BSs of the system. EE is given by

\begin{equation}
EE_{\zeta}(\eta)=\frac {C_{\zeta}(B, \eta, \rho)}{P_{tot} (\eta)}. \label{eqn:SE_EE}
\end{equation}
The distributions of both SE and EE will be derived by using previously obtained distributions.

\subsubsection{Distribution of Spectral Efficiency}
\label{sec:Dist_SE}

Distribution of $SE$ can be obtained similarly to capacity per UE distribution as found in Section \ref{sec:Analitic}. Using the cumulative distribution of received power $P_r$ (\ref{eqn:C_Zeta}), for user type $\zeta$ and time sharing parameter $\eta$, the distribution of $SE$ can be obtained as given by

\begin{equation}	
	F_{SE}(s|\zeta, B, \eta) = F_{Pr}(\sigma^2_{N+I, \zeta} (2^{\frac {s} {\eta_{\zeta}} }-1)), \label{eqn:SE_Zeta}
\end{equation}

By using (\ref{eqn:SE_Zeta}), the CDF of $SE$ for any UE in the system can be written as

\begin{equation}	
	F_{SE}(s|B,\eta) = \displaystyle \sum _{\zeta=1}^{3} P(\zeta|B) F_{SE}(s|\zeta, B, \eta), \label{eqn:SE_Any_UE}
\end{equation}
where $P(\zeta|B)$ is the probability of being type $\zeta$ UE in the system for a bias value of $B$.

\subsubsection{Distribution of Energy Efficiency}
\label{sec:Dist_EE}

In order to obtain EE which is given by (\ref{eqn:SE_EE}), $P_{tot} (\eta)$, total power consumed by BSs should be calculated. Here, we use a Base Station power consumption model which is proposed in \cite{GAUER}. In \cite{GAUER}, the BS power consumption is modeled by a linear power model:

\begin{equation}
	P_{in}=\begin{cases} N_{TRX} P_0 + \Delta_p P_{out}, & \mbox{if } \mbox{0 $<$} P_{out} \mbox{$\leq P_{max}$} \\ N_{TRX} P_{sleep}, & \mbox{if } P_{out} \mbox{=0} \end{cases}  \label{eqn:P_BS_Cons},
\end{equation}
where, $P_{in}$ is the total power consumed, $N_{TRX}$ is the number of transceivers in BS, $P_0$ is the power consumption at the minimum non-zero output power, $\Delta_p$ is the slope of the load-dependent power consumption, $P_{out}$ is the output power which is limited by $P_{max}$ and $P_{sleep}$ is the sleep mode power consumption of BS. The values of the parameters of Macro and Micro BSs are listed in Table \ref{table:BS_Power_Parameters}.

\begin{table}
	\centering
	\caption{Base Station Power Consumption Parameters}
	\begin{tabular}{|c|c|c|c|c|}
		\hline BS Type & $N_{TRX}$ & $P_{0}$ & $\Delta_{p}$ & $P_{sleep}$ \\ \hline
		Macro & 6 & 130 & 4.7 & 75 \\ \hline
		Micro & 2 & 56 & 2.6 & 39  \\ \hline
		
	\end{tabular}
	\label{table:BS_Power_Parameters}
\end{table}

By considering the model given by (\ref{eqn:P_BS_Cons}), the total power consumed by BSs of the system is given by

\begin{equation}
	\begin{aligned}
		P_{tot}=(1-\eta) P_{in,Macro} (P_{out}=P_{t,Macro}) \\ 
		+ \eta P_{in,Macro} (P_{out}=0) \\ 
		+N_{MICRO}  P_{in,Micro} (P_{out}=P_{t,Micro}).
	\end{aligned}
	\label{eqn:P_tot_BS}
\end{equation}

In (\ref{eqn:P_tot_BS}), $P_{in,Macro} (P_{out}=P_{t,Macro})$ is the total power consumed by Macro BS when the output power is $P_{t,Macro}$ and similarly $P_{in,Micro} (P_{out}=P_{t,Micro})$ is the total power consumed by a Micro BS when the output power is set to be $P_{t,Micro}$.

Using distribution of received power per UE which is given by (\ref{eqn:C_Zeta}), the distribution of $EE$ can be derived as given by

\begin{equation}
	F_{EE}(e|\zeta, B, \eta, \rho) = F_{Pr}(\sigma^2_{N+I, \zeta} (2^{\frac {e P_{tot} (\eta) N_{BS, \zeta}(B)} {\eta_{\zeta}W\rho_{\zeta}} }-1)). \label{eqn:EE_Zeta}
\end{equation}

By using (\ref{eqn:EE_Zeta}), the CDF of $EE$ for any UE in the system can be expressed as

\begin{equation}	
	F_{EE}(e|\eta) = \displaystyle \sum _{\zeta=1}^{3} P(\zeta|B) F_{EE}(e|\zeta,B, \eta, \rho). \label{eqn:EE_Any_UE}
\end{equation}

\section{Numerical Results}
\label{sec:Results}

	\begin{figure*}
		\centering
		
		\subfloat[$B=10dB$]{\label{subfig:CDF_B10}
		\includegraphics[height=7cm]{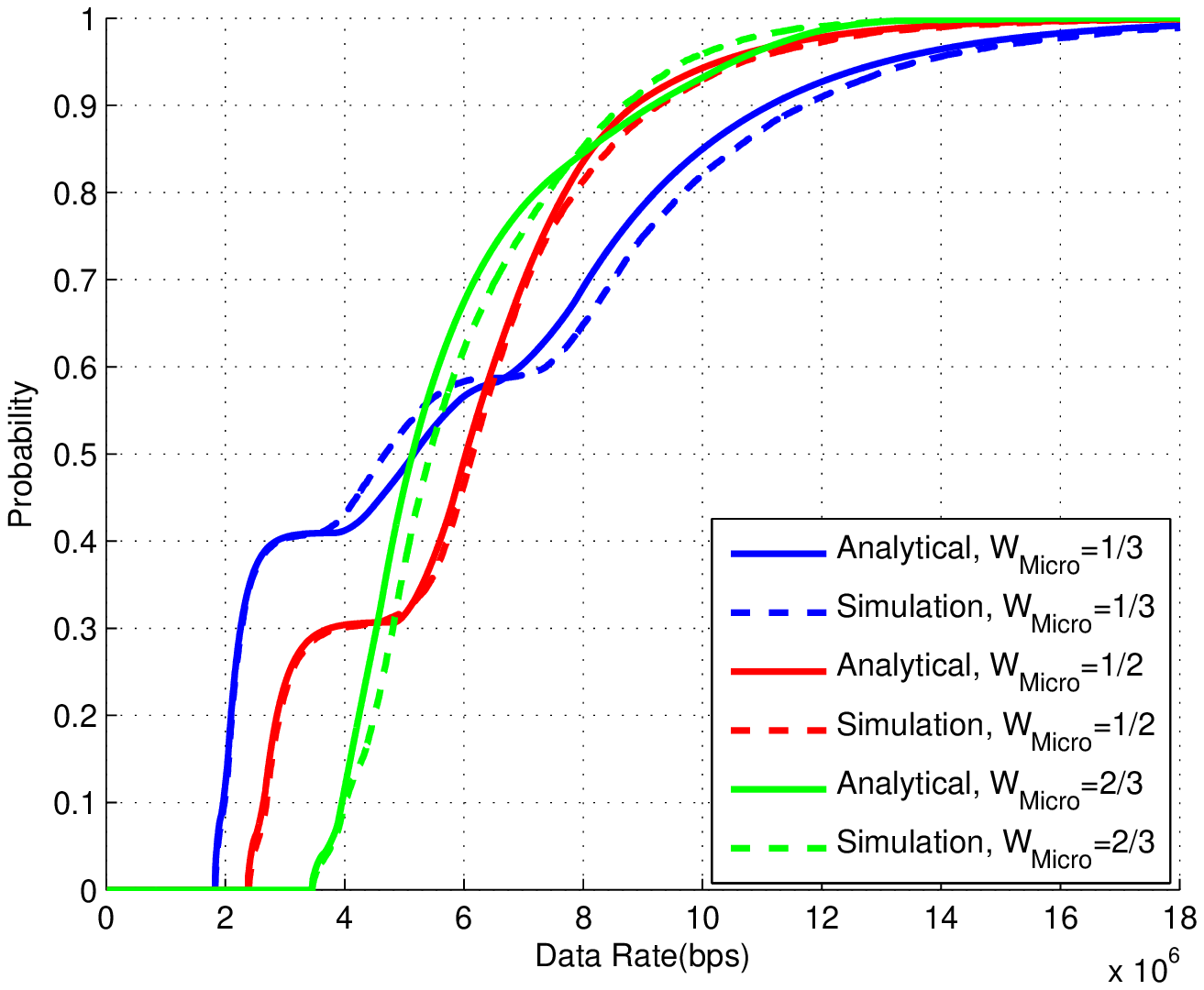}}
		\subfloat[$B=20dB$]{\label{subfig:CDF_B20}
		\includegraphics[height=7cm]{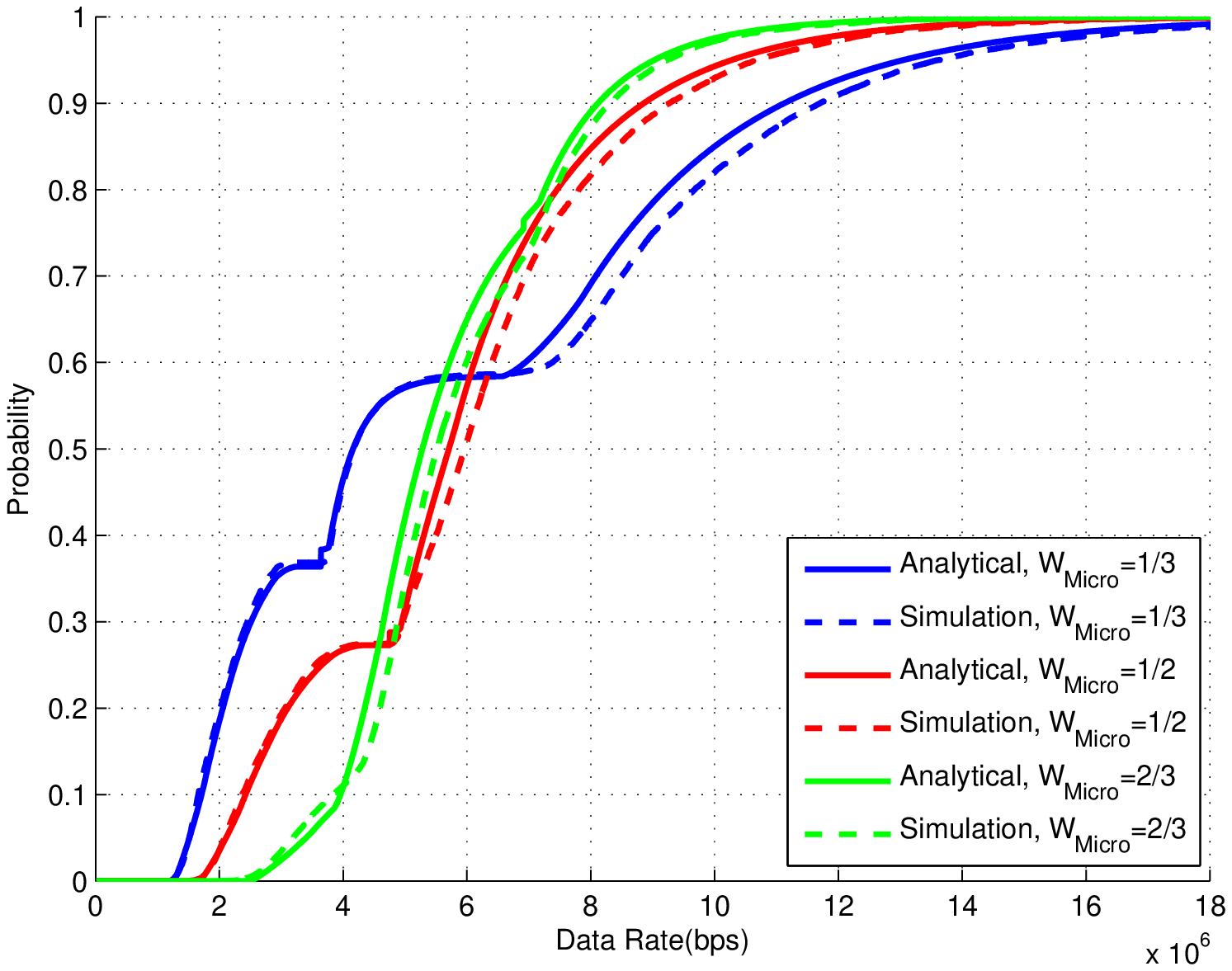}}
		\caption{CDF of Downlink Data Rate per User for $\eta = 0.2$, $\rho=0.5$} 
		\label{fig:CDF_10_20_dB}
	\end{figure*}

	\begin{figure*}
		\centering
		\subfloat[$B=10dB$]{\label{subfig:R10_3D_13_B10}
			\includegraphics[height=7cm]{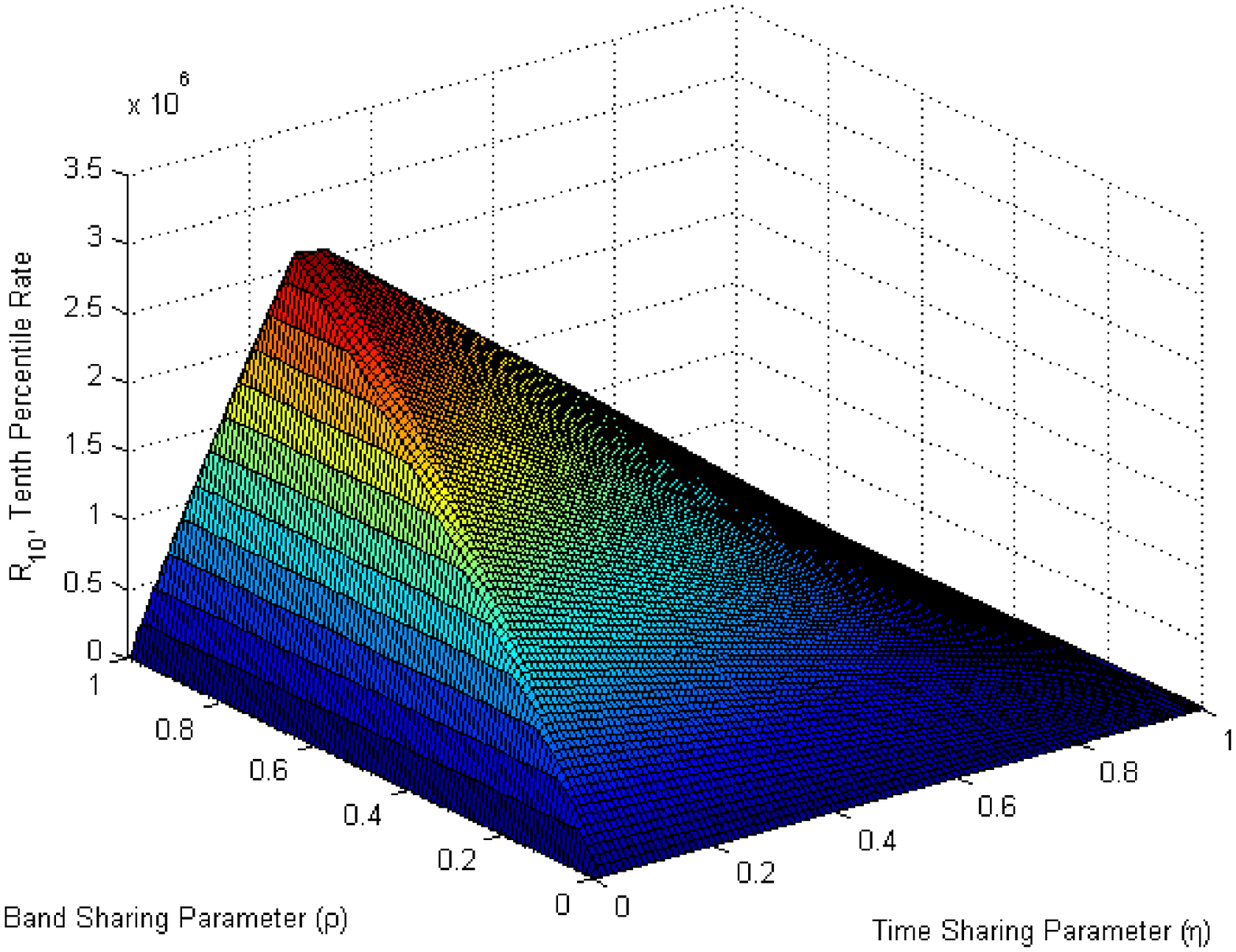}}
		\subfloat[$B=20dB$]{\label{subfig:R10_3D_13_B20}
			\includegraphics[height=7cm]{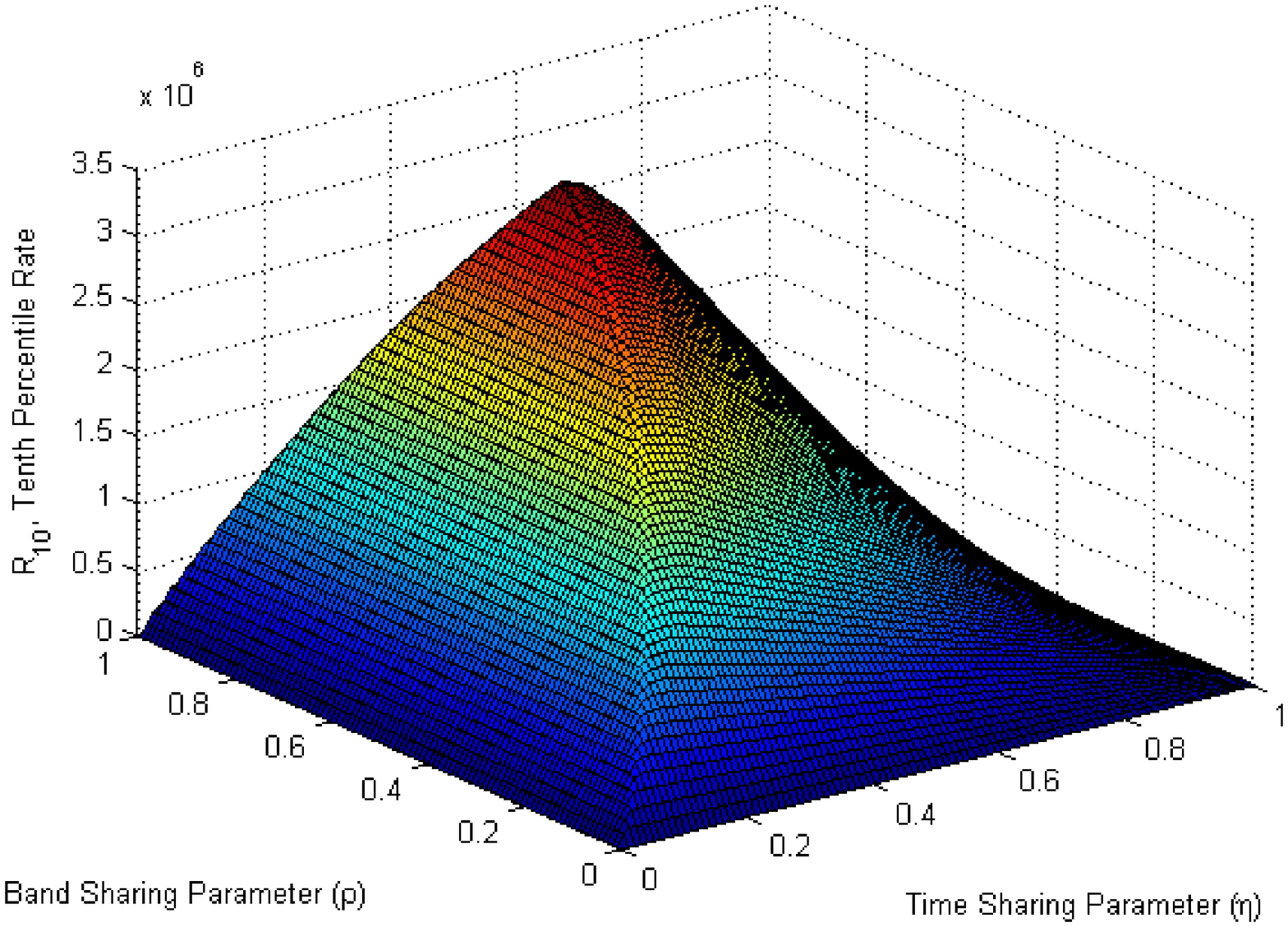}}
		\caption{Variation of $R_{10}$ as a function of $\rho$ and $\eta$ with $W_{Micro}=1/3$}
		\label{fig:R10_3D_W_1_3}
	\end{figure*}

	\begin{figure*}
		\centering
		\subfloat[$B=10dB$]{\label{subfig:R10_3D_12_B10}
		\includegraphics[height=7cm]{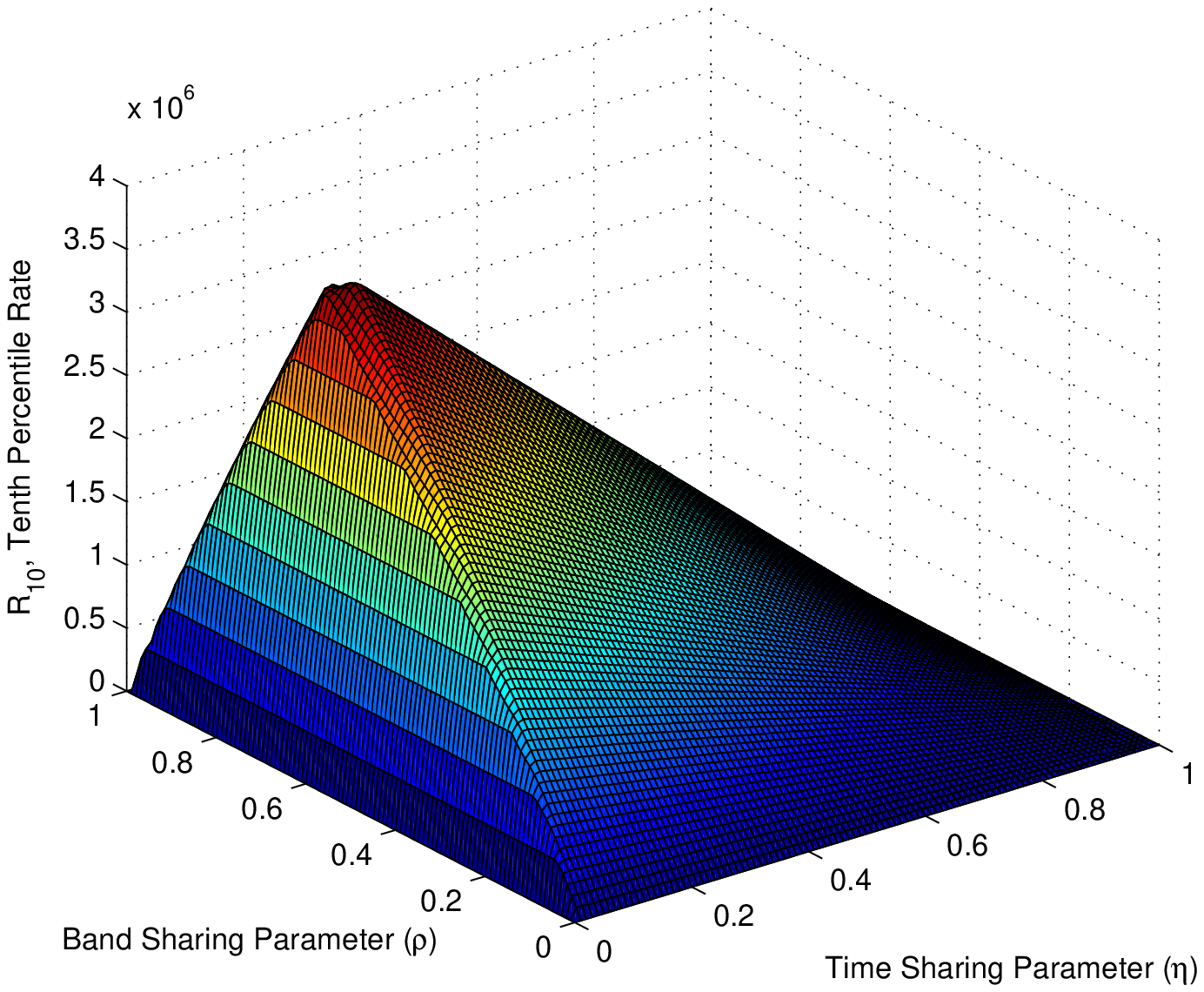}}
		\subfloat[$B=20dB$]{\label{subfig:R10_3D_12_B20}
		\includegraphics[height=7cm]{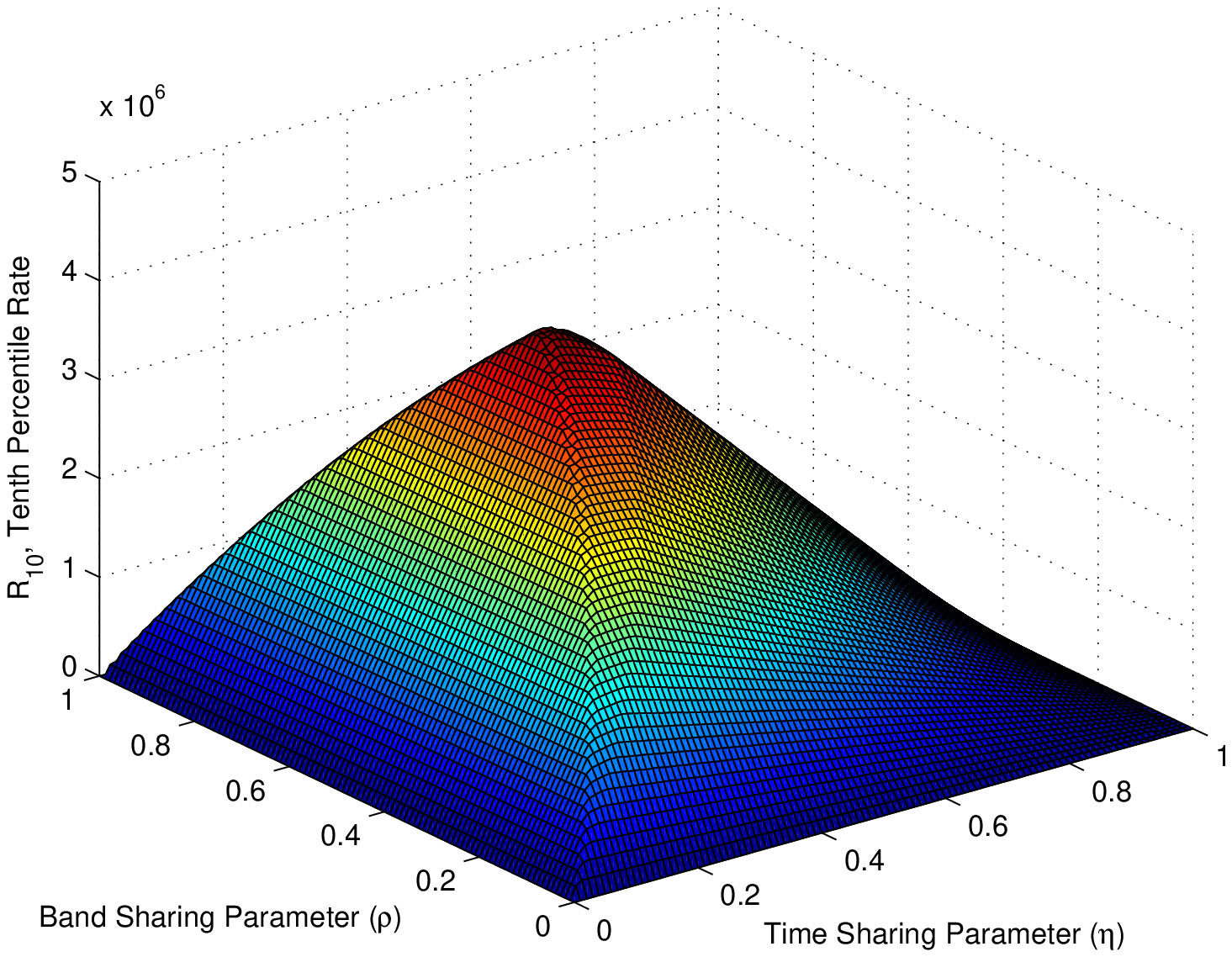}}
		\caption{Variation of $R_{10}$ as a function of $\rho$ and $\eta$ with $W_{Micro}=1/2$} 
		\label{fig:R10_3D_W_1_2}
	\end{figure*}

	\begin{figure*}
		\centering
		\subfloat[$B=10dB$]{\label{subfig:R10_3D_23_B10}
		\includegraphics[height=7cm]{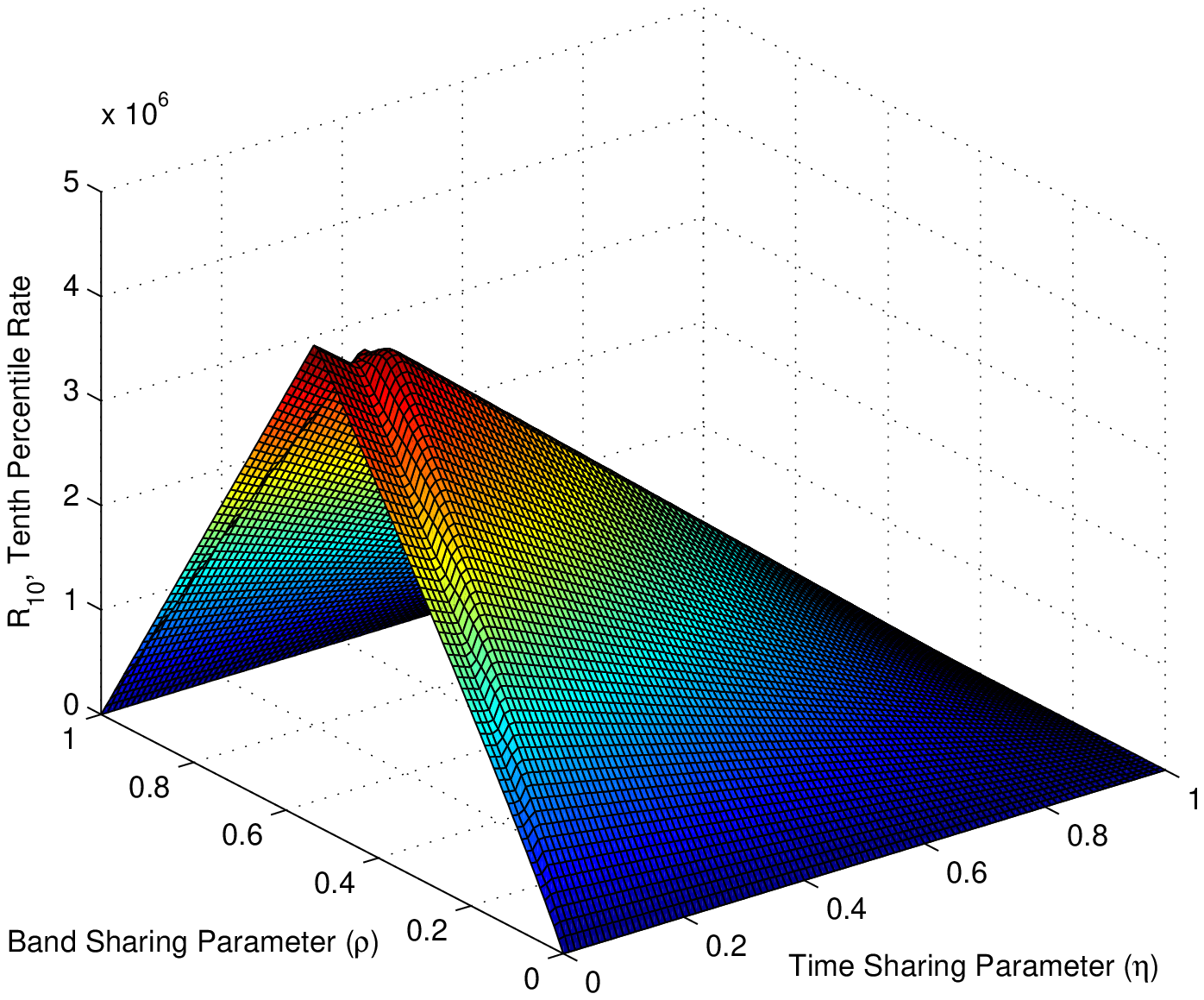}}
		\subfloat[$B=20dB$]{\label{subfig:R10_3D_23_B20}
		\includegraphics[height=7cm]{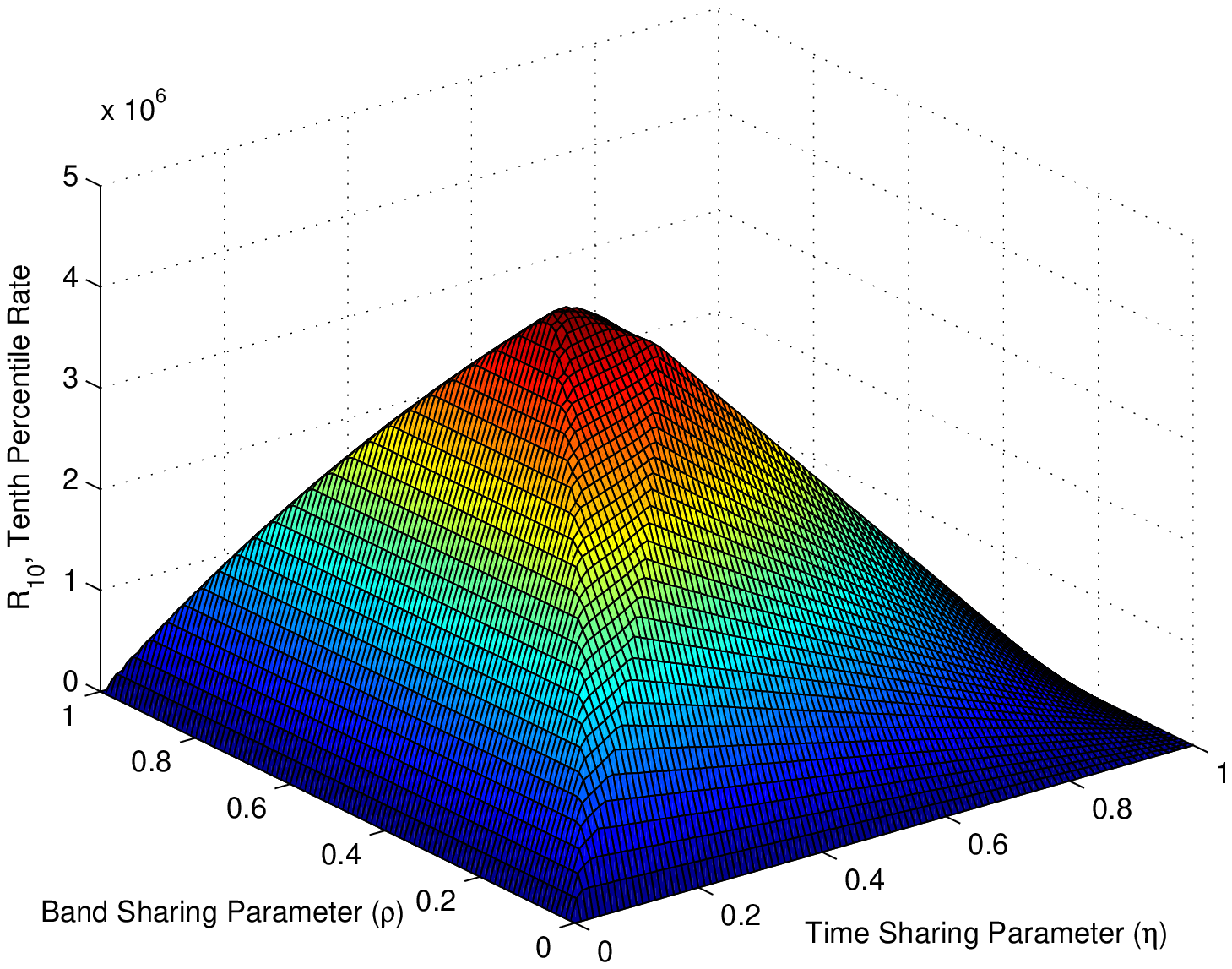}}
		\caption{Variation of $R_{10}$ as a function of $\rho$ and $\eta$ with $W_{Micro}=2/3$}  
		\label{fig:R10_3D_W_2_3}
	\end{figure*}

In this section, analytical results obtained for cumulative distribution of rate per UE will be compared first with the rate distribution obtained by simulations. The comparisons have been done for different bias ($B$) and resource allocation parameter ($\eta$, $\rho$) values and also for different UE distributions. Then, the analytical rate distribution has been employed in order to optimize system parameters $\eta$, $\rho$ and $B$ for different UE distributions. The optimizations have been done by considering tenth percentile rate $R_{10}$, which is the parameter we have selected as KPI in the system. Optimization of $R_{10}$ has also been done by simulations for comparison aspects. We have investigated the system in terms of Energy and Spectral efficiency. By using the $\rho$ values that maximizes $R_{10}$, we have obtained the variation of tenth percentile and median of EE and SE by both using analytical expression and doing simulations. The analytical and simulation results are obtained with system model parameter values which are listed in Table \ref{table:sim_pars}.

\begin{table}
	\centering
	\caption{Parameter values}
	\begin{tabular}{|c|c|}
		\hline  
		Parameter & Value  \\ \hline
		$P_{t,Macro}$ & $16$ dB \\ \hline
		$P_{t,Micro}$ & $-4$ dB \\ \hline
		$P_{noise}$ & $-173$ dBm/Hz \\ \hline
		$BS Noise Figure$ & $7$ dB \\ \hline
		$W$ & $100$ MHz \\ \hline
		$\eta$ & $0 \leq \eta \leq 1$ \\ \hline
		$\rho$ & $0 \leq \rho \leq 1$ \\ \hline
		$\alpha_{1}$ & $3.5$ \\ \hline
		$\alpha_{2}$ & $4$ \\ \hline
		$N_{UE}$ & 1000 \\ \hline
		$N_{MICRO}$ & 10 \\ \hline
		$N_{MACRO}$ & 1 \\ \hline
	\end{tabular}
	\label{table:sim_pars}
\end{table}

The cumulative probability distribution of rate per UE obtained by analytical formula and simulations are plotted in Figure \ref{fig:CDF_10_20_dB}. The CDFs are obtained for $\eta=0.2$, $\rho=0.5$, $B=10dB$ and $B=20dB$ and also for different UE distributions where $W_{micro}=\frac{1}{3}, \frac{1}{2}, \frac{2}{3}$. The goodness of fit of CDFs obtained by analytical approximation and simulations are compared by using Kolmogorov-Smirnov test \cite{KS_TEST}. Given the analytical distribution, this test shows whether random variables obtained empirically are distributed with the analytical distribution or not for a given level of significance value. In order to test the CDF obtained by analytical approximation, 100 UEs among 1000 UEs are randomly selected and the average level of significance between empirical and analytical distribution is calculated for 400 trials. The average significance values obtained for different UE distributions and $B$ values are given in Table \ref{subtable:KS_R_Significance} for rate distribution. Table \ref{subtable:KS_R_Ratio} shows the ratio of the KS tests passed for a significance value of $0.05$, which is a typical significance value for KS test. Tables \ref{subtable:KS_R_Significance}, \ref{subtable:KS_R_Ratio} and Figure \ref{fig:CDF_10_20_dB} show us that the CDFs obtained by analytical approximation and simulations are very close to each other. Consequently, we have concluded that analytical CDF approximation can be used for optimization of the system in terms of $R_{10}$.

\begin{table}
	\centering
	\caption{Kolmogorov-Smirnov Test Results for CDF of $R$} 
	\subfloat[Average Significance Values]{\label{subtable:KS_R_Significance}
		\begin{tabular}{|c|c|c|c|}
			\hline \backslashbox{$B$(dB)}{$W_{Micro}$} & 1/3 & 1/2 & 2/3 \\ \hline
			10 & 0.3320 & 0.4606 & 0.1609   \\ \hline
			20 & 0.4149 & 0.2740 & 0.2300 \\ \hline
		\end{tabular}}
	
	\subfloat[Pecentage of Kolmogorov-Smirnov Tests Passed for $P_{sig}=0.05$]{\label{subtable:KS_R_Ratio}
		\begin{tabular}{|c|c|c|c|}
			\hline \backslashbox{$B$(dB)}{$W_{Micro}$} & 1/3 & 1/2 & 2/3 \\ \hline
			10 & 0.8783 & 0.9033 & 0.5400   \\ \hline
			20 & 0.9117 & 0.7650 & 0.6867 \\ \hline
		\end{tabular}}
	\label{table:KS_R}
\end{table}

In order to find optimal values of system parameters, we have plotted the variation of $R_{10}$ with respect to $\rho$ and $\eta$ for two different bias values, $B=10dB$ and $B=20dB$ and for 3 different UE distributions. The variation of $R_{10}$ are presented in Figures \ref{fig:R10_3D_W_1_3}, \ref{fig:R10_3D_W_1_2} and \ref{fig:R10_3D_W_2_3}.

\begin{figure*}
	\centering
	\subfloat[$B=10dB$]{\label{subfig:Rate_10thPerc_B10}
		\includegraphics[height=7cm]{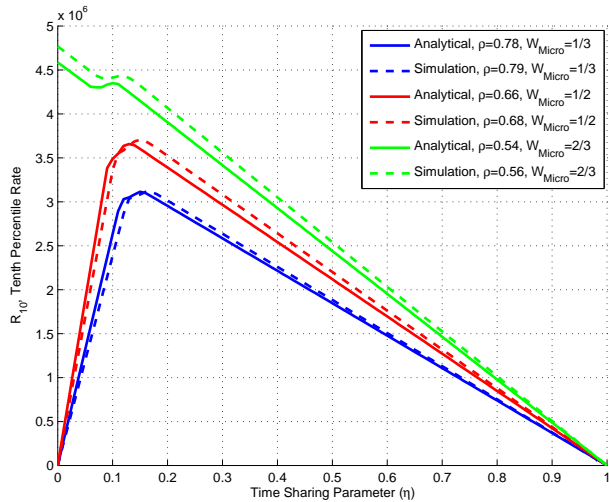}}
	\subfloat[$B=20dB$]{\label{subfig:Rate_10thPerc_B20}
		\includegraphics[height=7cm]{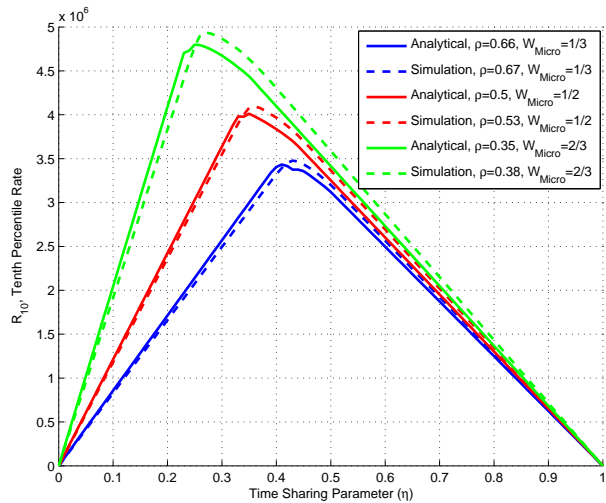}}
	\caption{$10^{th}$ Percentile Downlink Data Rate for varying $\eta$ and $W_{Micro}$} 
	\label{fig:Rate_10thPerc_t3}
\end{figure*}

Figure \ref{fig:Rate_10thPerc_t3} shows the cross-sections of Figures \ref{fig:R10_3D_W_1_3}, \ref{fig:R10_3D_W_1_2} and \ref{fig:R10_3D_W_2_3} which are plotted for $\rho$ values that maximizes $R_{10}$. Tables \ref{subtable:B10_Res_Comparison} and \ref{subtable:B20_Res_Comparison} show the optimal $\eta$ and $\rho$ values and maximum $R_{10}$ values that are obtained by simulations and using analytical CDF expression for different UE distributions.

By examining Figures \ref{fig:Rate_10thPerc_t3} and Tables \ref{subtable:B10_Res_Comparison} and \ref{subtable:B20_Res_Comparison}, it can be observed that analytically obtained results are very close to simulation results. It can also be concluded that the optimal $\rho$ value decreases and optimal $\eta$ value increases with increasing values of $B$. Optimal $\rho$ value also decreases with increasing $W_{micro}$, which says that as number of direct Micro UEs increases, larger portion of the bandwidth should be given to Micro BSs compared to small $W_{micro}$ values.

\begin{table}
	\centering
		\caption{The Comparison of Optimal Parameter Values and $R_{10}$} 
	\subfloat[$B=10dB$]{\label{subtable:B10_Res_Comparison}
	\begin{tabular}{|c|c|c|c|c|c|c|}
		\hline $W_{micro}$ & $\eta $ (S) & $\eta $ (A) & $\rho$ (S) &  $\rho$ (A)  & $R_{10}$ (S) & $R_{10}$ (A)   \\ \hline
		$1/3$ & 0.17& 0.15&0.79 &0.78 &3.115 & 3.113 \\ \hline
		$1/2$ & 0.15& 0.13& 0.68& 0.66 &3.704 & 3.657  \\ \hline
		$2/3$ & 0& 0& 0.56& 0.54& 4.768& 4.585  \\ \hline	
	\end{tabular}}

	\subfloat[$B=20dB$]{\label{subtable:B20_Res_Comparison}
		\begin{tabular}{|c|c|c|c|c|c|c|}
		\hline $W_{micro}$ & $\eta $ (S) & $\eta $ (A) & $\rho$ (S) &  $\rho$ (A)  &$R_{10}$ (S) & $R_{10}$ (A)   \\ \hline
		$1/3$ &0.43 &0.41 & 0.67& 0.66& 3.433 & 3.416 \\ \hline
		$1/2$ &0.36 & 0.35 & 0.53& 0.50& 4.095& 4.011  \\ \hline
		$2/3$ & 0.27& 0.25 & 0.38& 0.35 & 4.799 & 4.786  \\ \hline	
		\end{tabular}}
		\label{table:Res_Comparison}
\end{table}

Tables \ref{table:KS_SE} and \ref{table:KS_EE} show the results of KS test when it is applied to the distributions of SE and EE. Examining these results, it can be concluded that the analytical distributions obtained can be used for further optimization of the system in terms of Spectral and Energy Efficiency.

By examining Tables \ref{table:KS_R}, \ref{table:Res_Comparison}, \ref{table:KS_SE} and \ref{table:KS_EE}, it can be concluded that the accuracy of the analytical model decreases with increasing $W_{Micro}$ and $B$ values. The reason behind this is the approximations done to simplify base station coverage models for different type of UEs.

\begin{table}
	\centering
	\caption{Kolmogorov-Smirnov Test Results for CDF of $SE$} 
	\subfloat[Average Significance Values]{\label{subtable:KS_SE_Significance}
	\begin{tabular}{|c|c|c|c|}
		\hline \backslashbox{$B$(dB)}{$W_{Micro}$} & 1/3 & 1/2 & 2/3 \\ \hline
		10 & 0.4153 & 0.3296 & 0.2001   \\ \hline
		20 & 0.3935 & 0.2968 & 0.2163 \\ \hline
	\end{tabular}}
	
	\subfloat[Pecentage of Kolmogorov-Smirnov Tests Passed for $P_{sig}=0.05$]{\label{subtable:KS_SE_Ratio}
	\begin{tabular}{|c|c|c|c|}
		\hline \backslashbox{$B$(dB)}{$W_{Micro}$} & 1/3 & 1/2 & 2/3 \\ \hline
		10 & 0.9233 & 0.8350 & 0.6683   \\ \hline
		20 & 0.8967 & 0.7950 & 0.6733 \\ \hline
	\end{tabular}}
	\label{table:KS_SE}
\end{table}

\begin{table}
	\centering
	\caption{Kolmogorov-Smirnov Test Results for CDF of $EE$} 
	\subfloat[Average Significance Values]{\label{subtable:KS_EE_Significance}
	\begin{tabular}{|c|c|c|c|}
		\hline \backslashbox{$B$(dB)}{$W_{Micro}$} & 1/3 & 1/2 & 2/3 \\ \hline
		10 & 0.4055 & 0.3606 & 0.1642   \\ \hline
		20 & 0.4144 & 0.2705 & 0.2410 \\ \hline
	\end{tabular}}
	
	\subfloat[Pecentage of Kolmogorov-Smirnov Tests Passed for $P_{sig}=0.05$]{\label{subtable:KS_EE_Ratio}
	\begin{tabular}{|c|c|c|c|}
		\hline \backslashbox{$B$(dB)}{$W_{Micro}$} & 1/3 & 1/2 & 2/3 \\ \hline
		10 & 0.9117 & 0.8383 & 0.5400   \\ \hline
		20 & 0.9067 & 0.7600 & 0.7000 \\ \hline
	\end{tabular}}
	\label{table:KS_EE}
\end{table}

\begin{figure*}
	\centering
	%

  \subfloat[]{\label{subfig:SE_B10_20} \includegraphics[height=10.5cm]{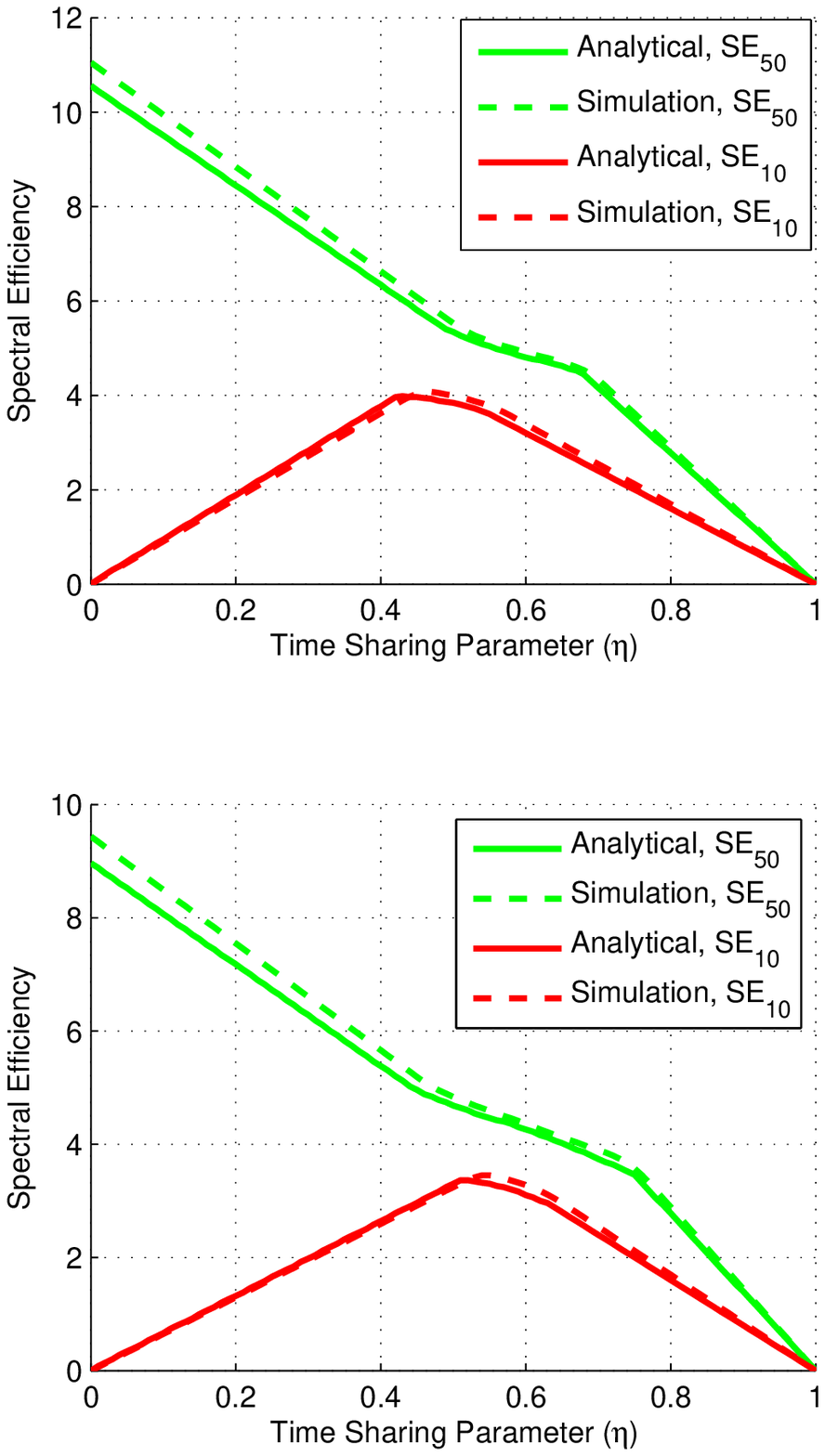}}
\subfloat[]{\label{subfig:EE_B10_20} \includegraphics[height=10.5cm]{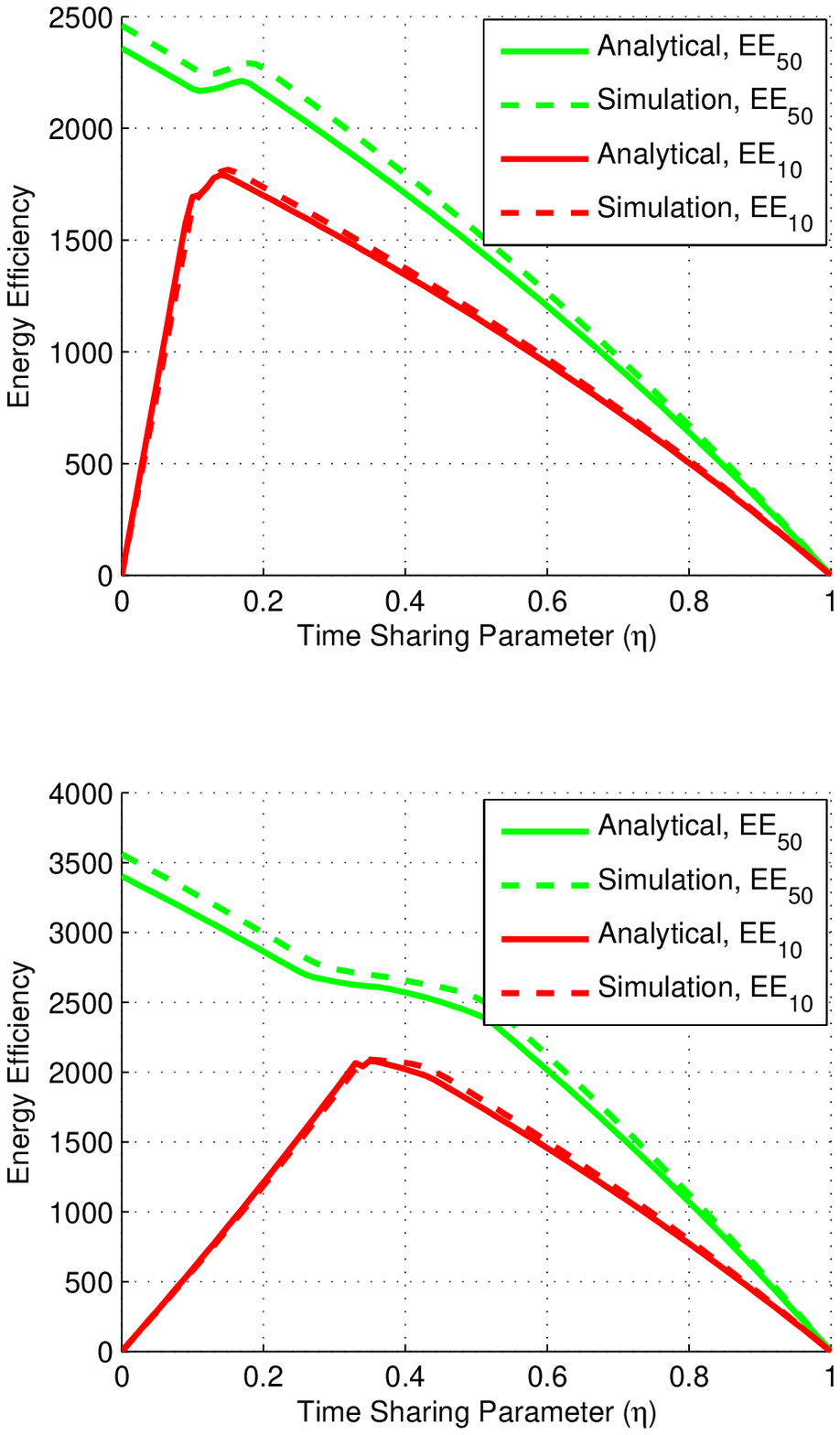}}
\subfloat[]{\label{subfig:theta_B10_20} \includegraphics[height=10.5cm]{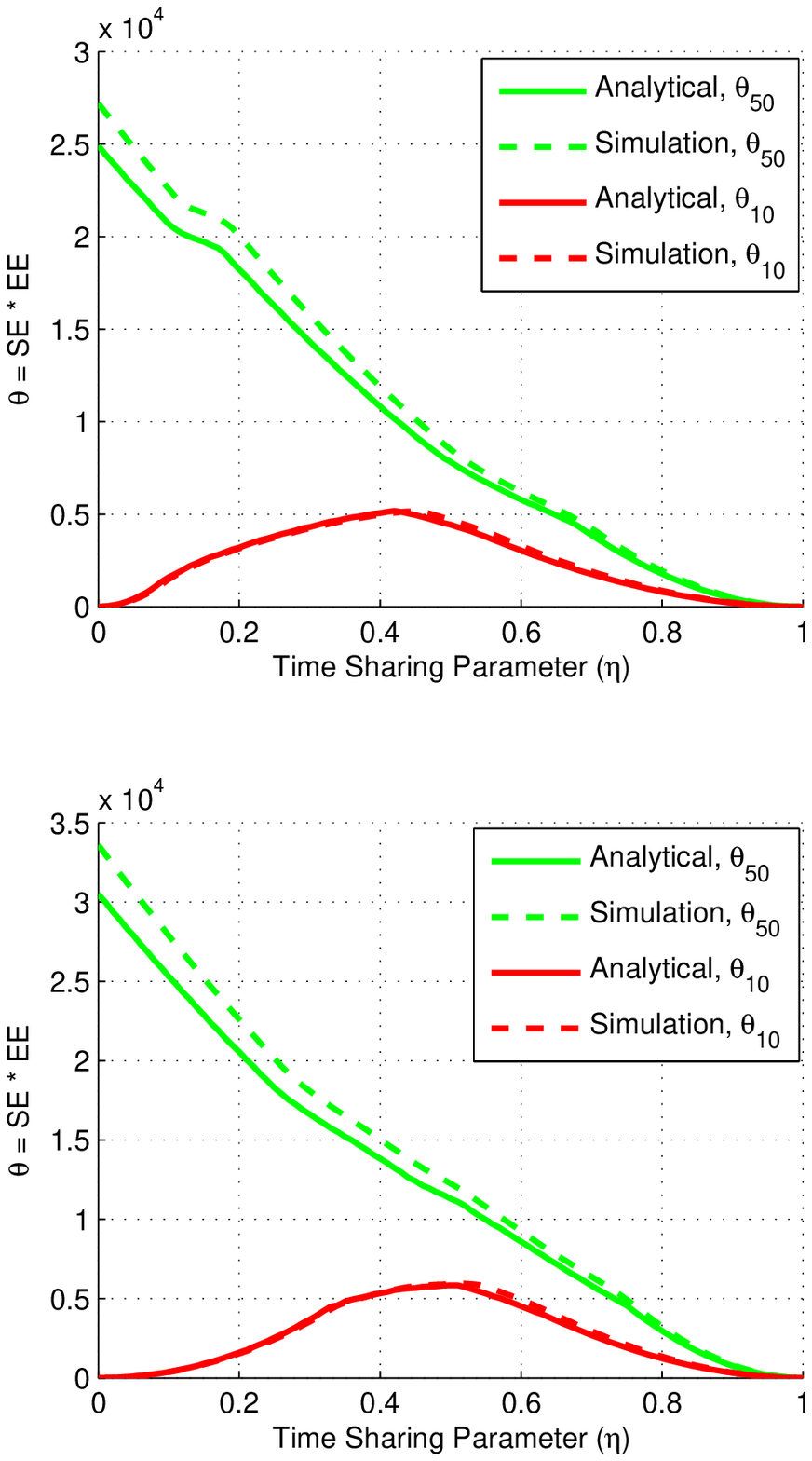}}
	\caption{Variation of $SE$, $EE$ and $\theta$ as a function of $\eta$ for $\rho=0.5$, $B=10dB$ (top), $B=20dB$ (bottom) and $W_{Micro}=1/2$} 
	\label{fig:SE_EE_Results}
\end{figure*}

\begin{figure}
			\centering
			\includegraphics[scale=0.6]{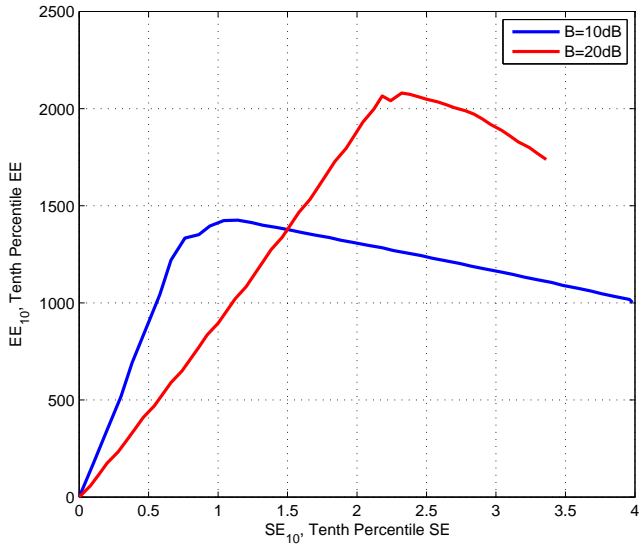}
			\caption{$SE_{10}$ vs $EE_{10}$ for $B=10dB, 20dB$ and $W_{Micro}=1/2$}
			\label{fig:SE10vsEE10_Results}
		\end{figure}

Figures \ref{subfig:SE_B10_20}, \ref{subfig:EE_B10_20} and \ref{subfig:theta_B10_20} shows the variation of tenth percentile and median of $SE$, $EE$, $\theta$ with varying $\eta$ and $B$ values, respectively.  By inspecting Figures \ref{subfig:SE_B10_20} and \ref{subfig:EE_B10_20}, it can be observed that median SE, and EE decays nearly linearly as $\eta$ increases. This result says us that Macro and Direct Micro UEs constitute more than half of the UEs consequently, as more resources are given to CRE UEs, less data rate, SE and EE are experienced by Macro and Direct Micro UEs. However, this is not the case if $SE_{10}$ and $EE_{10}$ are considered. By observing Figures \ref{subfig:SE_B10_20} and \ref{subfig:EE_B10_20}, it can be concluded that $SE_{10}$ and $EE_{10}$ are concave down functions of $\eta$ and $SE_{10}$, $EE_{10}$ are maximized at different values of $\eta$. When $EE_{10}$ is considered, the $\eta$ value that maximizes $EE_{10}$ is very close to the value that is optimal for $R_{10}$. This is because of the fact that $\eta$ is the parameter that controls $P_{tot}$. However, variation of $R_{10}$ with $\eta$ is much more faster that the variation of $P_{tot}$ with $\eta$. Therefore the variation of $P_{tot}$ with respect to $R_{10}$ can be assumed as constant. If we analyze the system in terms of $SE_{10}$, the situation is different. The $\eta$ value which maximizes $R_{10}$ and $EE_{10}$ does not maximize $SE_{10}$. As $\eta$ increases, the rate of CRE UEs increases however the SE of CRE UEs are still small compared to SE of Macro and Direct Micro UEs. Therefore larger $\eta$ is needed to increase $SE_{10}$. This result also exhibits the SE and EE trade-off in our Heterogeneous Network System model as also shown in Figure \ref{fig:SE10vsEE10_Results}. By inspecting Figures \ref{subfig:SE_B10_20}, \ref{subfig:EE_B10_20} and \ref{subfig:theta_B10_20}, it can also be concluded that the $\eta$ values maximizing $SE_{10}$, $EE_{10}$ and $\theta_{10}$ increase with $B$.

In order to jointly optimize SE and EE, two joint metrics $\theta_{10}$ and $\theta_{50}$ which are defined as in (\ref{eqn:SE_EE_10prod}) and (\ref{eqn:SE_EE_50prod}) are proposed.

\begin{equation}
	\theta_{10}=SE_{10} EE_{10} \label {eqn:SE_EE_10prod},
\end{equation}

\begin{equation}
\theta_{50}=SE_{50} EE_{50}, \label {eqn:SE_EE_50prod}
\end{equation}

Figure \ref{subfig:theta_B10_20} shows the variation of $\theta_{10}$ and $\theta_{50}$ with $\eta$. As observed from the figure, it can be concluded that the joint metric $\theta_{10}$ is maximized near the value that is optimal for $SE_{10}$ whereas the joint metric $\theta_{50}$ monotonically decreases with increasing $\eta$. This is due to the fact that $\theta_{50}$ covers the half of the UEs with small $\theta$ values. As $\eta$ increases, the resources given to Macro and Direct Micro UEs decreases, so $\theta_{50}$ decreases. From this observation, it can concluded that $\theta_{50}$ does not give any information about the variation of $\theta$ values of the UEs with smaller $\theta$ values. Therefore $\theta_{10}$ is a better metric if fairness is considered.

\section{Conclusion}
\label{sec:Conclusion}

In this paper, we have analyzed a Heterogeneous Network with cell-edge located small cells. In the system, there is one Macro and 10 Micro BSs and there are three types of UEs which are Macro, Direct Micro and CRE UEs.

By using a Time/Frequency resource allocation scheme that gives resources to different type of UEs orthogonally, we have obtained the CDFs of user rate, SE and EE by using a geometrical approach. We have compared these CDFs with CDFs obtained by extensive simulations. Our results show that the analytically derived CDFs are very close to the CDFs obtained by simulations. We have selected tenth percentile rate ($R_{10}$),  SE ($SE_{10}$) and EE ($EE_{10}$) as Key Parameter Indicators (KPIs) since these parameters inherently consider the fairness of the system and we have used the derived CDF expressions to optimize the system resource allocation parameters $\eta$, $\rho$ to maximize these KPIs. Our results show that the system is optimized around nearly same resource allocation parameters if $R_{10}$ and $EE_{10}$ are considered. However, larger $\eta$ values are needed to maximize $SE_{10}$, where  $R_{10}$ and $EE_{10}$ values degrade. This demonstrates Energy Efficiency and Spectral Efficiency trade-off in HetNet system under consideration. When SE and EE are jointly optimized by considering the proposed $\theta$ parameter, we observe that optimum value of $\eta$ is closer to the value of $\eta$ which maximizes SE. This shows that SE is more critical in the SE-EE trade-off.

\appendices
\section{Intersection Area Calculation of Two Circles}
\label{sec:TwoCircIntersection}
\begin{figure}
	\centering
	\includegraphics[scale=0.34]{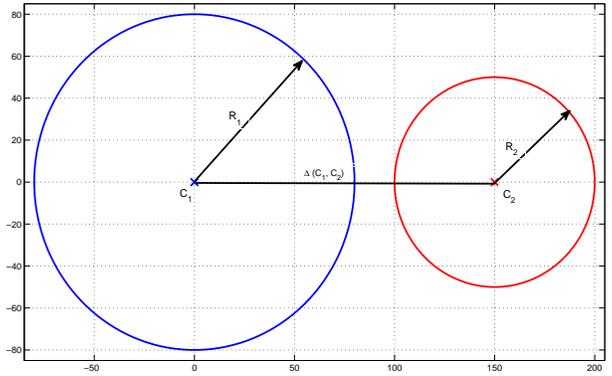}
	\caption{Intersection of two circles, Case-1} \label{fig:CC_Inters_1}
\end{figure}

\begin{figure}
	\centering
	\includegraphics[scale=0.36]{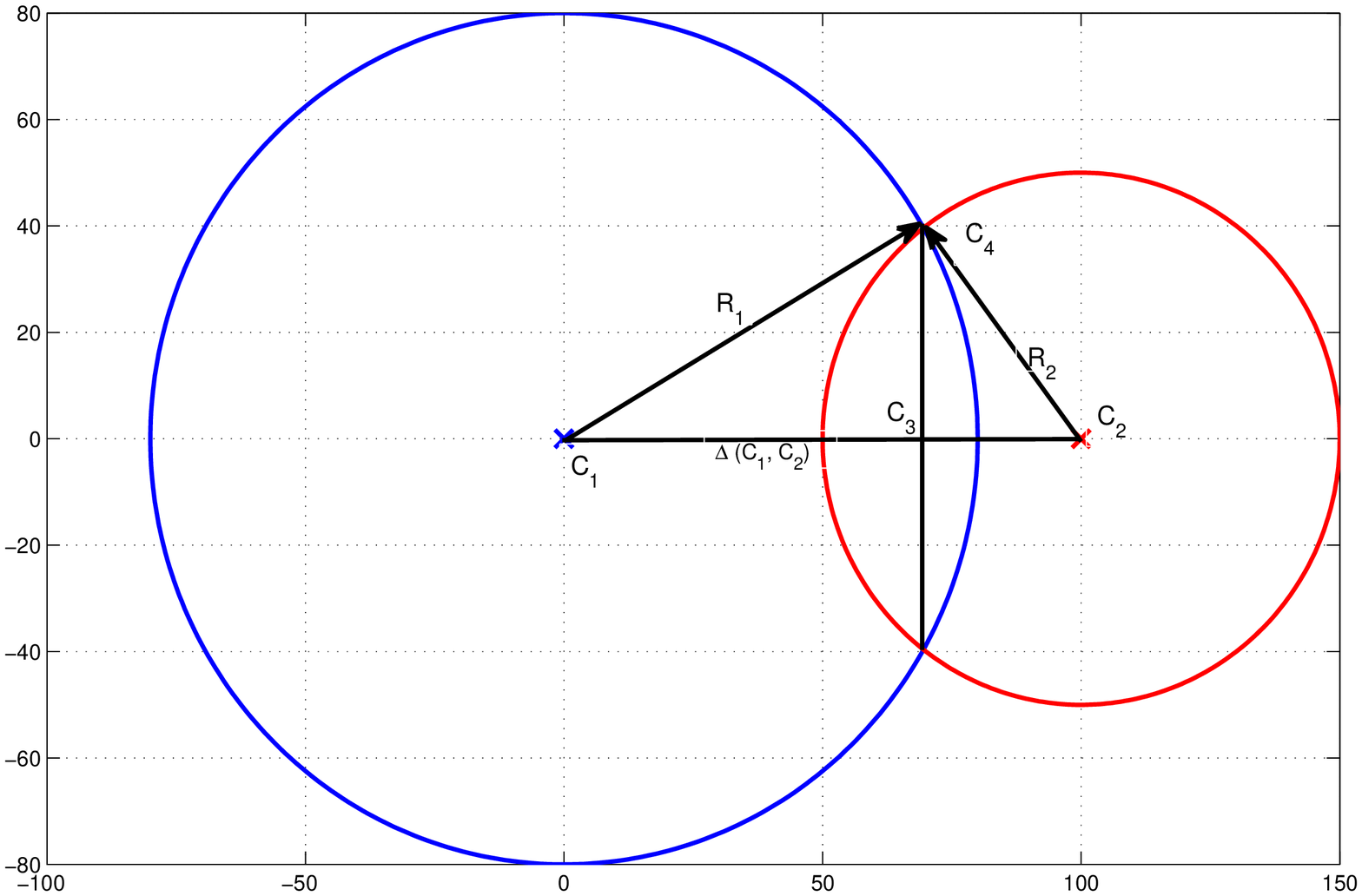}
	\caption{Intersection of two circles, Case-2} \label{fig:CC_Inters_2}
\end{figure}

\begin{figure}
	\centering
	\includegraphics[scale=0.5]{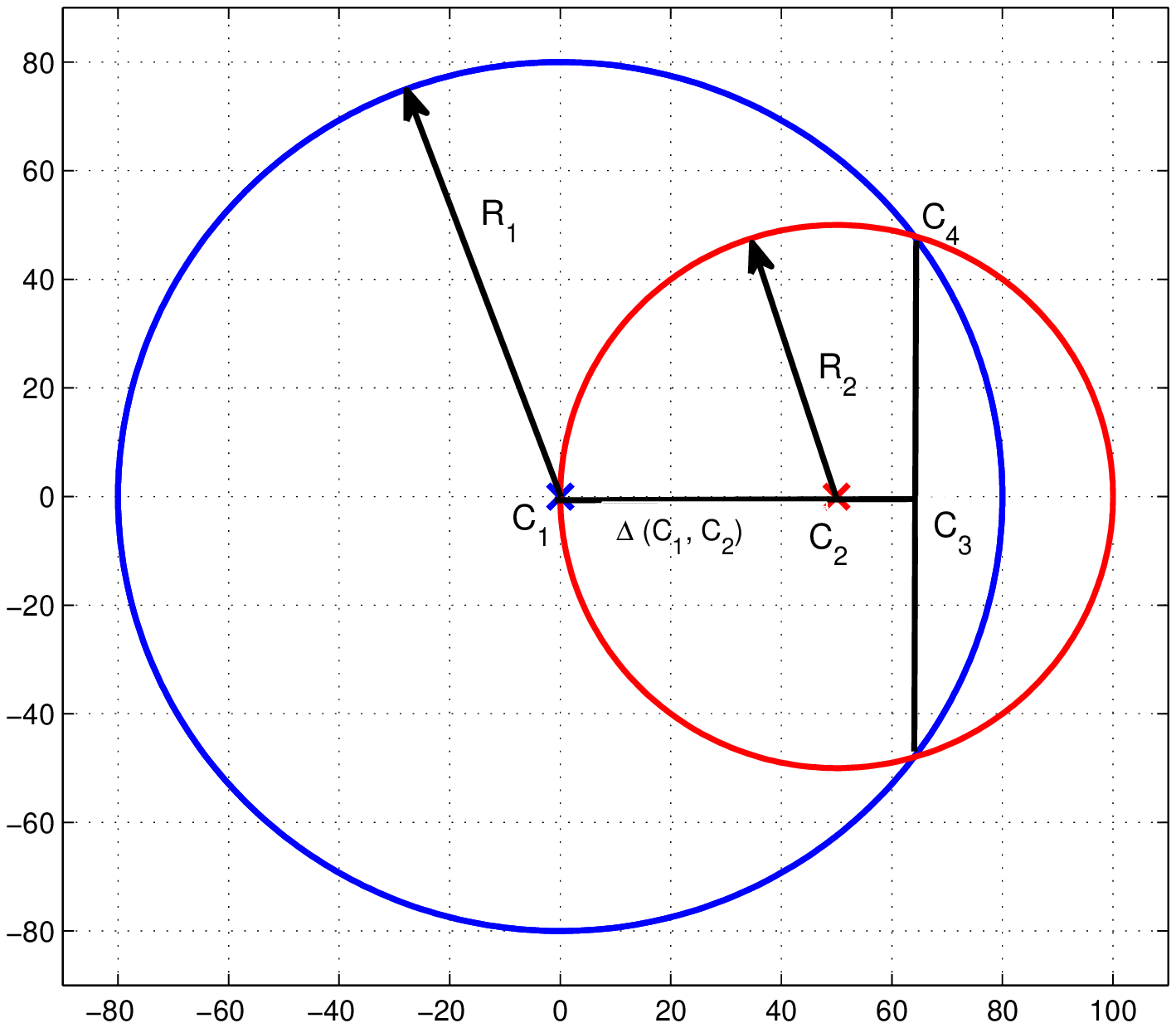}
	\caption{Intersection of two circles, Case-3} \label{fig:CC_Inters_3}
\end{figure}

\begin{figure}
	\centering
	\includegraphics[scale=0.5]{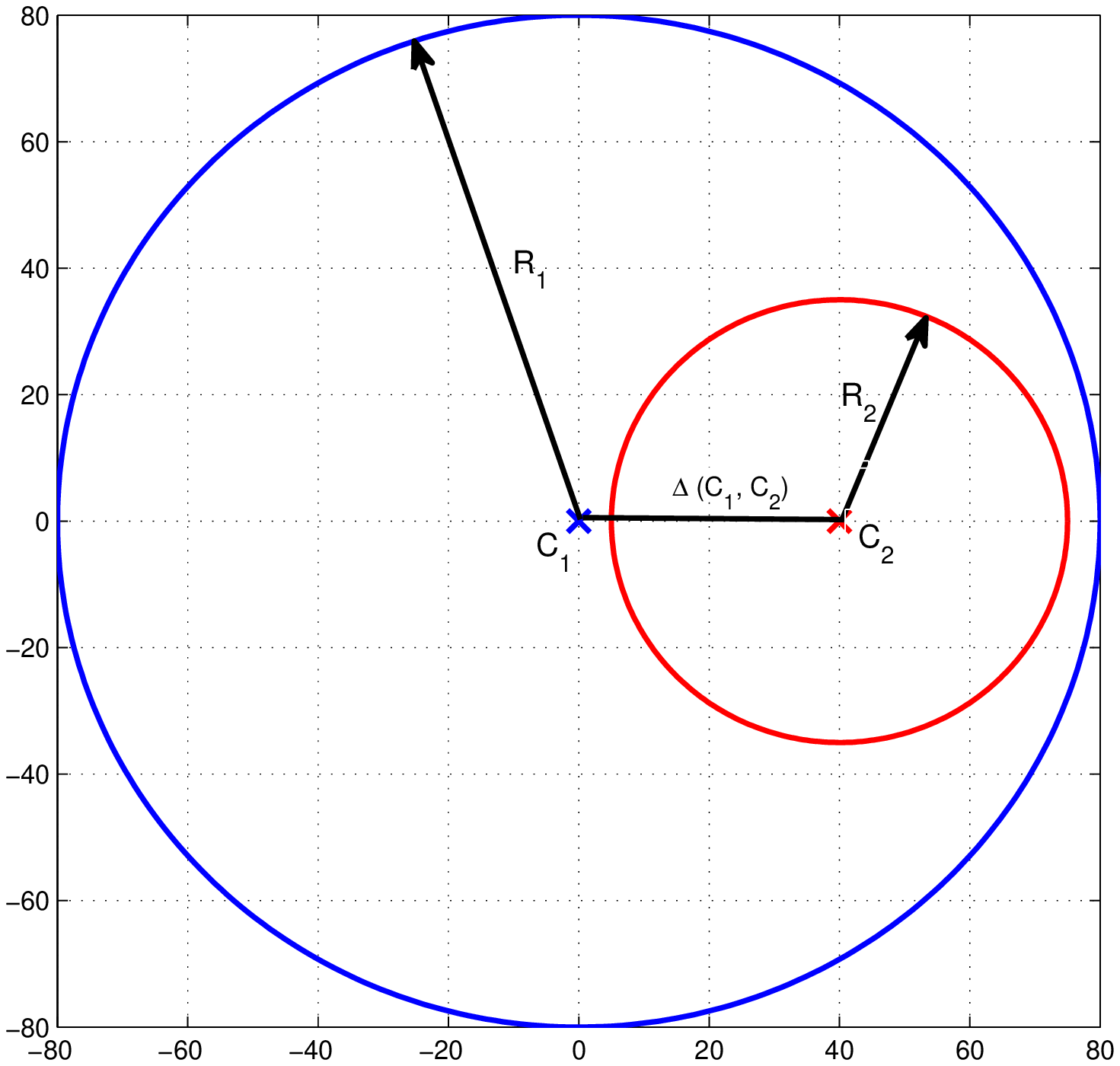}
	\caption{Intersection of two circles, Case-4} \label{fig:CC_Inters_4}
\end{figure}

Two circles with radius values $R_1$ and $R_2$ ($R_1>R_2$) may be located in four different positions to each other. These cases are shown in Figures \ref{fig:CC_Inters_1}-\ref{fig:CC_Inters_4}. In this section, formulas for the intersection area of these two circles for these four cases will be derived. The different cases in Figures \ref{fig:CC_Inters_1}-\ref{fig:CC_Inters_4} are defined as:

\begin{equation}
Case=\begin{cases}
0, & \mbox{$d > R_1+ R_2$}  \\
1, & \mbox{$R_1 < d < R_1+ R_2$} \\
2, & \mbox{$d < R_1 < 2R_2 $} \\
3 & \mbox{$d < R_1, R_1 > 2 R_2$} 
\end{cases}  \label{eqn:cases_all},
\end{equation}

In (\ref{eqn:cases_all}), $d$ is the distance between centers of the two circles and is defined by

\begin{equation}
d=|C_1-C_2|, \label{eqn:distance_centers}
\end{equation}
where $C_1$ and $C_2$ are the coordinates of centers of the two circles. 

(\ref{eqn:S_2_Inters}) gives the intersection area of two circles for different cases: $a$ and $b$ used in (\ref{eqn:S_2_Inters}) are distances between points $C_2,C_3$ and $C_3,C_4$, respectively. Points $C_2$ , $C_3$ and $C_4$ are shown in Figures \ref{fig:CC_Inters_2} and \ref{fig:CC_Inters_3}.

\begin{equation}
S_{intersect}=\begin{cases}
0 , & \mbox{Case 1}  \\

cos^{-1}(\frac {a}{R_1})R_1^2-ab\\+cos^{-1}(\frac{d-a}{R_2})R_2^2-b(d-a),& \mbox{Case 2, Case 3} \\
\pi R_2^2, & \mbox{Case 4} \\
\end{cases}  \label{eqn:S_2_Inters}
\end{equation}

where

\begin{equation}
a=\frac {-R_2^2+R_1^2+d^2}{2d}, \label{eqn:eqn_for_a}
\end{equation}

\begin{equation}
b=\sqrt{R_1^2-a^2}, \label{eqn:eqn_for_b}
\end{equation}

\section{Intersection Area Calculation of Three Circles}
\label{sec:ThreeCircIntersection}
\begin{figure}
	\centering
	\includegraphics[scale=0.37]{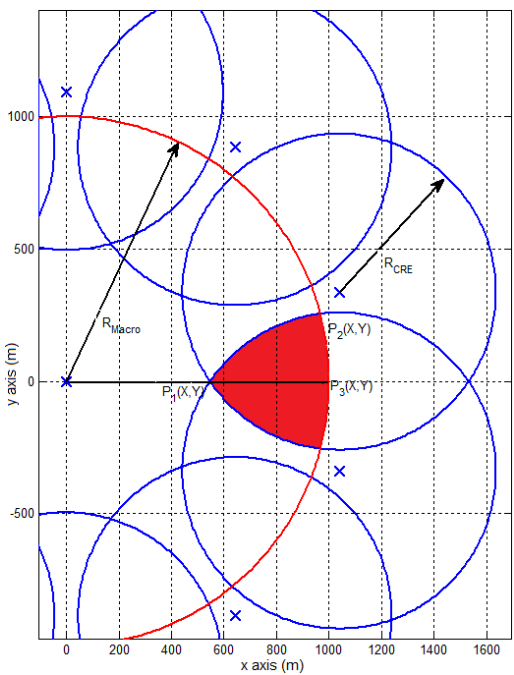}
	\caption{Intersection Area of 3 Circles} \label{fig:CC_Inters_3Circ}
\end{figure}

In order to properly calculate the region where CRE UEs are located, intersection area of three circles should be calculated. According to our system model, the red colored area in Fig. \ref{fig:CC_Inters_3Circ} should be obtained. Using definitions shown in Fig. \ref{fig:CC_Inters_3Circ}, the red area $S_{Int,3}$ can be expressed as

\begin{equation}
	S_{Int,3}=2(S_1+S_2)
	\label{eqn:Area_Sall}
\end{equation}
where

\begin{equation}
S_1=\int_{P_{1,x}}^{P_{2,x}} \sqrt{-(x-C_{1,x})^2+R_{CRE}^2}+ C_{1,y} dx 
\label{eqn:Area_S1},
\end{equation}

\begin{equation}
S_2=\int_{P_{2,x}}^{P_{3,x}} \sqrt{-(x-C_{2,x})^2+R_{Macro}^2}+ C_{2,y} dx
\label{eqn:Area_S2}
\end{equation}

In (\ref{eqn:Area_S1}), (\ref{eqn:Area_S2}), $P_{1,x}$, $P_{2,x}$, $P_{3,x}$ are the $x$ coordinates of the points $P_1(X,Y)$, $P_2(X,Y)$ and $P_3(X,Y)$, respectively. $P_1(X,Y)$ is the point where two adjacent CRE circle intersect, $P_2(X,Y)$ is the point where Macro circle and CRE circle intersects, $P_3(X,Y)$ is the point where $x$ axes and Macro circle intersects. $C_{1,x}$, $C_{1,y}$, $C_{2,x}$ and $C_{2,y}$ are the $x$ and $y$ coordinates of the centers of Macro and CRE circles respectively. $R_{Macro}$ and $R_{CRE}$ are the radius values of Macro and CRE circles. All these points and variables are shown in Figure \ref{fig:CC_Inters_3Circ}.

\ifCLASSOPTIONcaptionsoff
  \newpage
\fi



\bibliographystyle{IEEEtran}
\bibliography{biblio}
%

%

%





\end{document}